\documentclass[aps,prb,twocolumn,reprint,superscriptaddress
]{revtex4-2}

\usepackage{physics}
\usepackage[caption=false]{subfig}
\usepackage{graphicx}
\usepackage{amsmath,amssymb}
\usepackage[normalem]{ulem}
\usepackage{bm}
\usepackage{upgreek}

\usepackage{xcolor}

\usepackage{hyperref}
\hypersetup{
	citecolor = blue,
	colorlinks = true,
	urlcolor = blue
}

\begin{document}

\title{Microwave response of fractional quantum Hall droplets with quasiparticle tunneling}

\author{Fumihiro Murabayashi}
\email{fumihiro@issp.u-tokyo.ac.jp}
\affiliation{Institute for Solid State Physics, University of Tokyo, Kashiwa 277-8581, Japan}
\author{Ryotaro Sano}
\affiliation{Institute for Solid State Physics, University of Tokyo, Kashiwa 277-8581, Japan}
\author{Flavio Ronetti}
\affiliation{Aix Marseille Univ, Universit\'e de Toulon, CNRS, CPT, Marseille, France}
\author{J\'er\^ome Rech}
\affiliation{Aix Marseille Univ, Universit\'e de Toulon, CNRS, CPT,  Marseille, France}
\author{Thierry Martin}
\affiliation{Aix Marseille Univ, Universit\'e de Toulon, CNRS, CPT,  Marseille, France}
\author{Thibaut Jonckheere}
\affiliation{Aix Marseille Univ, Universit\'e de Toulon, CNRS, CPT,  Marseille, France}
\author{Takeo Kato}
\affiliation{Institute for Solid State Physics, University of Tokyo, Kashiwa 277-8581, Japan}
\email{kato@issp.u-tokyo.ac.jp}

\begin{abstract}
We theoretically study microwave absorption spectroscopy of fractional quantum Hall droplets in the presence of quasiparticle tunneling across a quantum point contact. This contact-free probe provides access to collective edge dynamics beyond conventional transport measurements. We develop a nonperturbative path-integral Monte Carlo approach that enables computation of the frequency-dependent response at finite temperature and for arbitrary droplet geometries, and benchmark the method against analytical results in the weak-tunneling regime. We find that tunneling produces measurable shifts and broadening of resonance peaks, with systematic dependence on tunneling strength and device geometry. Such shifts and broadenings are not obtained in perturbative treatments acting directly on the response function, but emerge when interaction-kernel effects are properly incorporated. Our results indicate experimentally accessible signatures of edge-mode interference and tunneling-induced renormalization of collective excitations, and support the use of microwave spectroscopy as a quantitative probe of quasiparticle dynamics in mesoscopic quantum Hall structures.
\end{abstract}

\maketitle

\section{Introduction} 
\label{sec:introduction}

Since its discovery, the fractional quantum Hall effect (FQHE) has had a profound impact
on condensed matter physics~\cite{Tsui1982PRL}.
Its essence lies in the formation of an incompressible quantum liquid driven by strong electron-electron interactions, and the remarkable agreement between theory and experiment, most notably exemplified by the Laughlin wave function~\cite{Laughlin83}, has greatly advanced our understanding of strongly correlated electronic phases.
Moreover, the FQHE represents a clear example of topological order that cannot be characterized by any local order parameter, and it has played a decisive role in establishing fundamental concepts in modern topological materials science.

In recent years, renewed attention has been directed toward quasiparticle excitations in fractional quantum Hall (FQH) systems.
For the Laughlin state realized at filling $\nu = 1/(2m+1)$ ($m \in \mathbb{Z}$), characterized by a single chiral edge mode encircling the bulk~\cite{Halperin82}, the elementary excitations carry fractional charge. A prominent example is the $\nu=1/3$ state, in which the quasiparticle contributing to edge transport has been experimentally observed to possess charge $e/3$, as revealed by shot-noise measurements~\cite{Saminadayar1997PRL,dePicciotto1997Nat}.
These quasiparticles are known to be Abelian anyons exhibiting exotic fractional statistics, whereby the many-body wave function acquires a phase factor upon exchanging two particles~\cite{Wilczek82,Arovas84,Nayak08,Stern2008,Feldman2021}.
Thanks to significant advances in nanofabrication and measurement technology, this fractional statistic has recently been directly verified through Fabry–P\'{e}rot interferometry and nonequilibrium noise measurements~\cite{Nakamura2020,Bartolomei2020Sci,Carrega2021}.
Motivated by these achievements, both theoretical and experimental studies on anyonic statistics in FQH systems have intensified rapidly relying either on interferometry~\cite{Nakamura2023PRX,Kundu2023Nat,Willett2023PRX,Batra2023,Werkmeister2024NatComm,Werkmeister2025Science} or time-domain braiding~\cite{Rosenow2016PRL,Han2016NatComm,Mora2022arXiv,Morel2022PRB,Lee2022Nat,Ruelle2023PRX,Glidic2023PRX,Jonckheere2023PRL,Schiller2023PRL,Lee2023Nat,Iyer2024PRL,Thamm2024PRL,Ronetti2025PRL,Flavio2025PRB,Ruelle2025Science}.

However, authors of most previous studies have focused on transport properties such as current and nonequilibrium noise in devices where a quantum point contact (QPC) or an interferometer is fabricated in the edge channel~\cite{Saleur02,Glattli05}.
Such setups necessarily require attaching source and drain electrodes, making it impossible to avoid the influence of the measurement apparatus itself, including contact resistance and potential fluctuations near the electrodes, on the intrinsic edge structure.
Against this background, probing edge excitations via microwave absorption in an isolated quantum Hall droplet~\cite{Volkov1988JETP,Andrei1988SurfSci,Talyanskii1990SurfSci} provides an experimental window fundamentally distinct from conventional transport measurements~\cite{Ashoori1994PRB,Zhitenev1994PRB,Kumada2011PRB,Petkovic2013PRL,Hashisaka2013PRB,Kumada2014PRL,Endo2018JPSJ}.
Related developments in quantum Hall plasmonic and microwave architectures have further highlighted the potential of edge excitations as a resource for spectroscopy and device functionalities~\cite{Mahoney2017PRX,Bosco2019,Frigerio2024CommPhys}.
Related spectroscopic and contactless approaches have also been discussed for probing quasiparticle dynamics and ac response in quantum Hall systems~\cite{Ferraro2014,Drichko2014}.
The setup proposed in Ref.~\cite{Cano2013PRB} capacitively couples a mesoscopic quantum Hall droplet to a microwave waveguide, allowing direct access to the excitation spectrum of edge modes without any Ohmic contacts.
This noninvasive feature effectively eliminates contact-induced disturbances, offering a decisive advantage over transport-based approaches.
Furthermore, this method offers several additional benefits that are difficult to realize in transport experiments.
First, the resonance frequencies of absorption peaks directly encode the propagation velocity of edge modes, while the number of peaks reflects the number of modes, allowing their physical properties to be determined in a highly direct manner.
Second, the set of resonance peaks shows a distinctive dependence on the shape of the droplet, offering a way to explore the geometry of the quantum Hall effect~\cite{Oblak2024PRX}.
Finally, when multiple edges are present, by simply varying the droplet size, one can effectively cross the equilibration length of the edge, thereby accessing both the equilibrated regime (single peak) and the nonequilibrated regime (multiple peaks) within the same device~\cite{Cano2013PRB}.
This stands in sharp contrast with transport measurements, where accessing different regimes typically requires altering the positions of electrical contacts~\cite{Kane95,Deviatov06}.

In addition, when a QPC is introduced, the absorption peak amplitude exhibits a characteristic interference effect that is linear in the tunneling amplitude, in contrast with the quadratic dependence observed in standard Fabry–P\'{e}rot interferometers.
This feature greatly enhances sensitivity to quasiparticle charge and statistics.
Moreover, contributions from the bulk have frequency and magnetic field dependences distinct from those of the edge~\cite{Mast85,Glattli85}, enabling high-precision extraction of pure edge physics from the absorption spectrum.

\begin{figure}[tb]
\begin{center}
\includegraphics[clip,width=6.0cm]{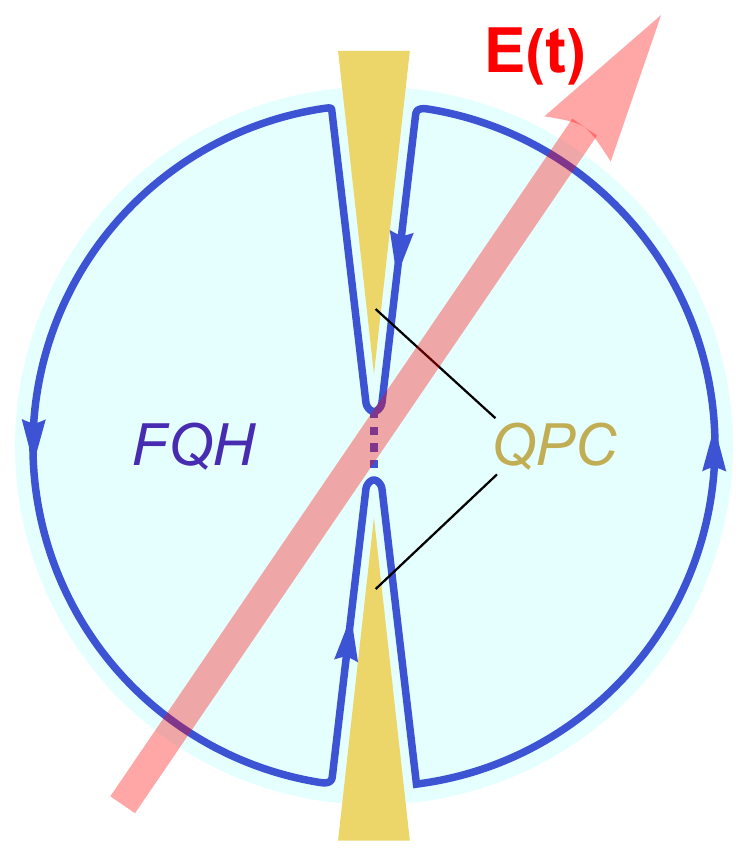}
\caption{Schematic view of the setup: a FQH droplet with a chiral edge state (blue line) containing a QPC (yellow), allowing tunneling of fractional quasiparticles. The droplet is driven by a time-varying microwave electric field (large red arrow), assumed spatially uniform over the droplet scale.}
\label{fig:setup}
\end{center}
\end{figure}

In this work, we investigate how quasiparticle tunneling through a QPC modifies the microwave absorption spectrum of a FQH droplet as shown in Fig.~\ref{fig:setup}.
We compute the finite-temperature response nonperturbatively using path-integral Monte Carlo (PIMC), which treats interactions, thermal fluctuations, and tunneling processes on equal footing and allows access to arbitrary droplet geometries. 
To gain analytical insight, we complement the numerics with perturbative calculations in the tunneling amplitude for an idealized circular geometry and compare the results. 
Benchmarking against the exact solution for an integer quantum Hall (IQH) droplet at $\nu=1$ demonstrates the consistency of both the analytical framework and numerical implementation.
This combined analysis shows that tunneling strength and temperature induce systematic resonance-peak shifts and linewidth broadening. 
The peak shifts arise beyond perturbative treatments acting directly on the response function~\cite{Cano2013PRB}, underscoring the importance of considering the interaction kernel in discussing the microwave response.
Throughout this paper, we set $\hbar = 1$.

\section{Model and Formulation} 
\label{sec:model}

\subsection{Bosonization}
\label{subsec:formulation eff action}

We consider a circular FQH state at a finite temperature $T$ (see Fig.~\ref{fig:setup}), hereafter referred to as a FQH droplet.
Such a droplet can be realized by applying a negative voltage to appropriately shaped gate electrodes in a two-dimensional electron system. 
Owing to the bulk energy gap, all low-energy excitations are confined to the boundary and appear as a single unidirectional chiral edge mode, as indicated by the blue line in Fig.~\ref{fig:setup}. 
Although this edge mode is described by the chiral Luttinger liquid theory~\cite{Chang2003RevModPhys,Wen1992IJMP}, careful treatment is required for finite-size systems~\cite{Michael1997PRB,vonDelft1998}.

We introduce a coordinate $s$ along the edge of the FQH droplet and define the edge-state region as $0 \le s < L$, where $L$ denotes the circumference. 
In the bosonization method, the normal-ordered electron density $\rho(s)$ is described with a bosonic field as
\begin{align}
    \rho(s) = \frac{1}{2\pi} \partial_s \varphi(s) ,
\end{align}
where $\varphi(s)$ is a bosonic field obeying the commutation relation
\begin{align}
[\varphi(s),\varphi(s')]= i\pi\nu \, \mathrm{sgn}(s-s'), \quad (0<s,s'<L).
\end{align}
For a finite system, this bosonic field can be decomposed into two contributions~\cite{Michael1997PRB,Loss1992PRL}
\begin{align}
    \varphi(s) &= \varphi_0(s)+\varphi_{\rm p}(s),\\
    \varphi_0(s)&=-\nu \vartheta + \frac{2\pi N s}{L},\\
    \varphi_{\rm p}(s) &= \sum_{k>0}\sqrt{\frac{2\pi \nu}{kL}}\left(b_ke^{iks} + b_{k}^{\dagger}e^{-iks}\right)e^{-k\alpha/2},
\end{align}
where $\varphi_0(s)$ and $\varphi_{\rm p}(s)$, respectively denote the zero mode and the periodic part, $N = \int_0^L ds \, \rho(s)$ is the charge operator, $\vartheta$ is its conjugate phase operator, and $b_{k}^{\dagger}$ ($b_{k}$) are bosonic creation (annihilation) operators. 
The summation runs over discrete wave numbers $k=2\pi n/L$, with $n$ a positive integer.
The canonical commutation relation $[\vartheta,N]=i$ holds by construction.

The quasiparticle annihilation operator is described with the bosonic field as~\cite{vonDelft1998}
\begin{align}
\psi_{\rm q}(s) &= \frac{1}{\sqrt{2\pi\alpha}} e^{-i\nu \vartheta}e^{2\pi i N s/L} e^{i\varphi_{\rm p}(s)}  \notag \\
&= \frac{1}{\sqrt{2\pi\alpha}} e^{i\varphi_0(s)+i\pi\nu s/L} e^{i\varphi_{\rm p}(s)} ,
\end{align}
 where $\alpha$ is a short-distance cutoff.
In the second equality, the operators have been rearranged using the Baker--Campbell--Hausdorff (BCH) formula~\cite{Magnus54}. 
The resulting $c$-number phase factor $e^{i\pi\nu s/L}$ originates from the noncommutativity of $\vartheta$ and $N$.

The Hamiltonian of the chiral edge mode is given by~\cite{vonDelft1998}
\begin{align}
H &= \frac{\pi v }{\nu L} \left[N^2 + (\phi_{\rm AB}/\pi + \nu)N \right] + \sum_{k>0} vk b_k^\dagger b_k ,
\end{align}
where $v$ is the velocity of the chiral mode.
Here we have introduced the Aharonov--Bohm (AB) phase $\phi_{\rm AB} = 2\pi \Phi / \Phi_0$, where $\Phi$ is the magnetic flux through the droplet, and $\Phi_0 = h/e^{*}$ is the flux quantum corresponding to the quasiparticle charge $e^{*}=\nu e$.
In the presence of the AB phase, the quasiparticle operator is replaced as
\begin{align}
\psi_{\rm q}(s) 
&= \frac{1}{\sqrt{2\pi\alpha}} e^{i\varphi_0(s)+i \phi_{\rm AB} s/L +i\pi\nu s/L} e^{i\varphi_{\rm p}(s)} .
\end{align}
In this work, we only consider the $N=0$ sector. This is justified when the effect of thermal fluctuations on the charge number $N$ can be neglected. To this end, we consider the low temperature regime ($k_{\rm B}T \ll 2\pi v/L$) and choose a phase $\phi_{\rm AB}$ close to $-\pi \nu$ where
the thermal fluctuations  of $N$ are minimized.
Furthermore, the phase operator $\vartheta$ can be dropped as we consider quasiparticle tunneling within a single edge mode, which does not modify the total number
of quasiparticles on the edge.
Therefore, the quasiparticle annihilation operator is simplified as
\begin{align}
\label{eq:fieldop}
\psi_{\rm q}(s) &= \frac{1}{\sqrt{2\pi\alpha}}
e^{i\varphi_{\rm p}(s)}.
\end{align}
In what follows, we denote $\varphi_{\rm p}(s)$ as $\varphi(s)$ to keep the notation as simple as possible.

\subsection{Effective action}

In the following, we derive the effective action by applying the path-integral formalism to the nonzero mode.
The imaginary-time action of the edge modes is given as follows~\cite{Moon1993PRL}
\begin{align}
    \label{eq:S_edge}
    S_{\rm{edge}}[\varphi] = \frac{1}{4\pi \nu}\int_0^L ds \int_0^{\beta} d\tau \, \partial_s \varphi \left(i\partial_{\tau} \varphi + v\partial_s \varphi \right) ,
\end{align}
where $L$ is the circumference of the circular FQH droplet, and $\beta = 1/k_{\rm B}T$ is the inverse temperature.

Next, we consider the effect of quasiparticle tunneling at a QPC. 
In this study, we focus on quasiparticle tunneling between positions $s_1$ and $s_2$ along the edge of the droplet.
For a tunneling amplitude $\lambda_r$, the action describing the quasiparticle tunneling is given by 
\begin{align}
    \label{eq:S tun quasi}
    S_{\rm{tun}}[\varphi] &= -\lambda_{r}\int_0^{\beta}d\tau\, \left[\psi^{\dagger}_{\rm{q}}(s_2,\tau)\psi_{\rm{q}}(s_1,\tau) + \rm H.c.\right] \notag\\
    &= -V \int_0^{\beta} d\tau\, \cos\left[\varphi (s_1,\tau) - \varphi (s_2, \tau) \right] ,
\end{align}
where in the last step we set $V=\lambda_r/\pi\alpha$. 
From this, the action of the system is $S_{\rm edge}[\varphi] + S_{\rm tun}[\varphi]$, and the partition function is expressed using path integrals as 
\begin{align}
    Z &\propto \int \mathcal{D}\varphi \, e^{-S_{\rm{edge}}[\varphi]-S_{\rm{tun}}[\varphi]} .
\end{align}
For the calculation of the response function, we employ the generating functional with an auxiliary real field $\lambda(\tau)$, which allows us to carry out the Gaussian integration everywhere except at the positions $s_1$ and $s_2$
\begin{align}
    \label{eq:gene functional}
    Z &\propto \int \mathcal{D}\varphi \, \mathcal{D}\lambda \, \mathcal{D}\phi\, e^{-S_{\rm{edge}}[\varphi]-S_{\rm{tun}}[\varphi]-S_{\lambda}[\varphi,\lambda,\phi]},
\end{align}
where we introduced the action involving the newly defined fields
\begin{align}
    \label{eq:S_lambda}
    S_{\lambda}[\varphi,\lambda,\phi] &= i\int_0^{\beta} d\tau \,\lambda(\tau)[\varphi(s_1,\tau)-\varphi(s_2,\tau) - \phi(\tau)] .
\end{align}
Here, $S_{\lambda}[\varphi,\lambda,\phi]$ enforces the constraint between the fields at $s_1$ and $s_2$.

After performing a Fourier transform, completing the square, and carrying out Gaussian integrations over the fields $\lambda$ and $\varphi$, the generating functional reduces to an action depending only on the local field $\phi(\tau)$, which corresponds to the effective action $S_{\rm eff}[\phi]$ of the system
\begin{align}
    \label{eq:gene eff}
    Z &\propto \int \mathcal{D}\phi\, e^{-S_{\rm{eff}}[\phi]}, \\
    \label{S_eff}
    S_{\rm{eff}}[\phi] &= S_0[\phi] + S_{\rm{tun}}[\phi] .
\end{align}
Here, $S_0[\phi]$ and $S_{\rm{tun}}[\phi]$ represent the contributions from the edge states and quasiparticle tunneling, respectively, and are given by
\begin{align}
    S_0[\phi] &=\sum_{\omega_n}\frac{\left|\phi_n\right|^2}{4\beta\Lambda_n} , 
    \label{eq:S0 result} \\
    S_{\rm{tun}}[\phi] &= -V \int_0^\beta d\tau \, \cos\left[\phi(\tau)\right] .
\end{align}
Here, $\phi_n$ is a Fourier transformation of $\phi(\tau)$
\begin{align}
    \label{eq:phiFourier}
    \phi(\tau) &= \frac{1}{\beta}\sum_{\omega_n}\phi_ne^{-i\omega_n\tau} ,
\end{align}
and $\omega_n \equiv 2\pi n/\beta$ denotes the bosonic Matsubara frequencies.
$\Lambda_n$ is a function dependent on $s_1$ and $s_2$, which is given by
\begin{align}
    \Lambda_n = \frac{ 2 \pi \nu}{\omega_n}\frac{\sinh\left(\frac{\omega_n\left|s_1-s_2\right|}{2v}\right)\sinh\left(\frac{\omega_n\left(L-\left|s_1-s_2\right|\right)}{2v}\right)}{\sinh\left(\frac{\omega_n L}{2v}\right)} .
    \label{eq:Lambda_n}
\end{align}
For a detailed derivation, see Appendix~\ref{app:effective_action}.
One readily sees from Eq.~\eqref{eq:S0 result} that $\Lambda_n$ is directly related to the propagator of the effective bosonic field $\phi$.

\subsection{Response function}
\label{subsec:formulation res}

To formulate the microwave response function of the droplet, we first define the polarization of the edge states as follows
\begin{align}
    \label{eq:polarization}
    \hat{P} = \int_0^L ds\, y(s) \rho(s) ,
\end{align}
where $y(s)$ denotes the $y$ coordinate of the edge of the droplet, along which the direction of the oscillating electric field is aligned.
We assume that the microwave has a sufficiently long wavelength so that the electric field remains constant over the scale of the droplet.

\begin{figure}[tb]
\begin{center}    
\includegraphics[clip,width=8.0cm]{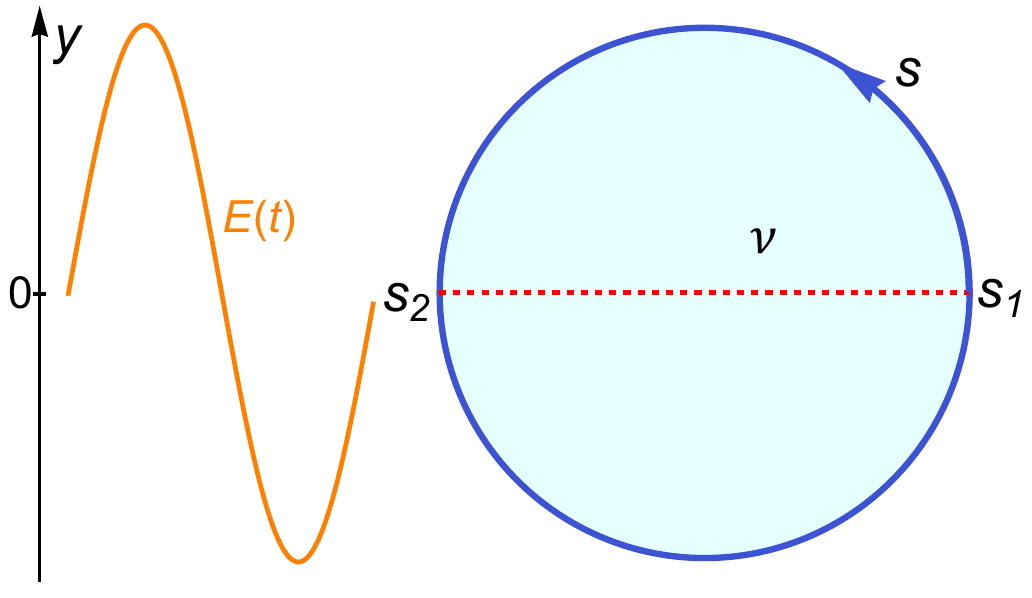}
\caption{Schematic of a quantum Hall droplet with circular geometry at filling $\nu$. 
The coordinate $y$ is defined parallel to the oscillating microwave electric field, while the coordinate $s$ parametrizes the circumference of the droplet. 
The points $s_{1}$ and $s_{2}$ denote the positions of the QPCs. 
Although a realistic droplet would exhibit constrictions at these locations, we consider an idealized setup in which the droplet remains circular and particle (or quasiparticle) tunneling occurs at $s_{1}$ and $s_{2}$.}
\label{fig:circular_model}
\end{center}
\end{figure}

In this work, we focus on an idealized setup in which the droplet remains circular and particle (or quasiparticle) tunneling occurs at $s_{1}$ and $s_{2}$, as shown in Fig.~\ref{fig:circular_model}, serving as a representative example. 
For a circular droplet with circumference $L$, $y(s)$ is given by
\begin{align}
    \label{circle y(x)}
    y(s) = \frac{L}{2\pi} \sin\!\left(\frac{2\pi s}{L}\right).
\end{align}
In the rest of this section, we formulate the microwave response for this setup, whereas the extension to different geometries can be carried out analogously and will be discussed in Sec.~\ref{sec:RealisticGeometry}.

The Hamiltonian describing the interaction between the microwave electric field and the edge states is then given by
\begin{align}
    \label{polarization hamiltonian}
    \hat{H}^{\rm{(ex)}}(t) &= -E(t)\hat{P}
    = -E(t)\int_0^L ds\, y(s) \rho(s) .
\end{align}
We define the expectation value of the polarization change due to the external field $E(t)$ as $\delta \langle \hat{P}(t)\rangle \equiv \langle \hat{P}(t)\rangle - \langle \hat{P} \rangle$. The Fourier transforms of the electric field and the polarization change are then given by
\begin{align}
    E(t) &= \int_{-\infty}^\infty \frac{d\omega}{2\pi} \, E(\omega) e^{-i\omega t}, \\
    \delta \langle \hat{P} (t) \rangle  &= \int_{-\infty}^\infty \frac{d\omega}{2\pi} \, \delta P(\omega) e^{-i\omega t}.
\end{align}
The linear response of the polarization to the electric field is summarized in the form $\delta P(\omega) = \epsilon^R(\omega) E(\omega)$ by linear response theory, where the susceptibility $\epsilon^R(\omega)$ is given by the Fourier transform of the retarded polarization correlation function $\epsilon^R(t) =i \Theta(t)\langle [ \hat{P}(t),\hat{P}(0)]\rangle$.
Here, $\hat{P}(t)$ denotes the time evolution of the polarization in the absence of the external field (in the Heisenberg picture). 
The susceptibility $\epsilon^R(\omega)$ can be calculated from the imaginary-time evolution of the polarization as follows
\begin{align}
    \epsilon^R(\omega) &= \epsilon(i\omega_n \rightarrow \omega + i\delta) , \\
    \label{eq:res fourier it to Mf}
    \epsilon(i\omega_n) &= \int_0^{\beta} d\tau \, \epsilon(\tau) e^{i\omega_n \tau} , \\
    \label{eq:res it def}
    \epsilon(\tau) & = \langle T_\tau \hat{P}(\tau) \hat{P}(0) \rangle .
\end{align}
Here, $\hat{P}(\tau)$ represents the imaginary-time evolution of the polarization in the absence of the external field, which can be expressed by replacing the charge density $\rho(s)$ in Eq.~(\ref{eq:polarization}) with its imaginary-time evolution $\rho(s,\tau)$. 
Because the microwave absorption intensity is directly related to ${\rm Im}\,\epsilon^R(\omega)$ through the fluctuation-dissipation theorem, we restrict our attention to this quantity in what follows.

From Eqs.~(\ref{eq:res fourier it to Mf}) and (\ref{eq:res it def}), the response function in the Matsubara representation is calculated as
\begin{align}
    \label{eq:permittivity}
    \epsilon(i\omega_n) &= \int_0^L ds ds' \, y(s) y(s') S_{\rho\rho}(s,s',i\omega_n) .
\end{align} 
Here, $S_{\rho\rho}(s,s',i\omega_n)$ is given as the Fourier transform of the imaginary-time density–density correlation function of the edge state $S_{\rho\rho}(s,s',\tau) \equiv \langle T_\tau \rho(s,\tau) \rho(s',0)\rangle$. The charge density $\rho(s,\tau)$ is rewritten with the bosonic field as $\rho(s,\tau) = (2\pi)^{-1}\partial_s \varphi(s,\tau)$.
To evaluate $S_{\rho\rho}(s,s',\tau)$, we introduce a source term coupled to an external field $\eta(s,\tau)$ into the generating functional defined in Sec.~\ref{subsec:formulation eff action}, as follows:
\begin{align}
    \label{eq:gene functional with eta}
    Z[\eta] &\propto \int \mathcal{D}\varphi \, \mathcal{D}\lambda \, \mathcal{D}\phi\, e^{-S_{\rm{edge}}[\varphi]-S_{\rm{tun}}[\varphi]-S_{\lambda}[\varphi,\lambda,\phi]-S_{\eta}[\varphi]}, \\
    \label{eq:S_eta}
    S_{\eta}[\varphi] &= \frac{1}{2\pi} \int_0^\beta d\tau \! \int_0^L ds \, \eta( s,\tau) \, \partial_s\varphi(s,\tau) . 
\end{align}
The correlation function can be calculated directly from the generating functional $Z[\eta]$, introduced in Eq.~\eqref{eq:gene functional with eta}, as
\begin{align}
    S_{\rho\rho}(s,s',\tau,\tau')
    &\equiv \frac{1}{4\pi^2}\langle T_\tau \partial_{s}\varphi(s,\tau)\partial_{s'}\varphi(s',\tau')\rangle \nonumber \\    
    &= \left. \frac{1}{Z[\eta]}\frac{\delta^2 Z[\eta]}{\delta \eta(s,\tau)\delta\eta(s',\tau')}\right|_{\eta = 0} .
    \label{eq:dens-correlate}
\end{align} 

\subsection{Response function in a circular droplet}
\label{subsec:response circular}

By functional differentiation of the resulting generating functional, we can obtain the explicit expression for the density correlation function
\begin{align}
    \label{eq:charge dens corr}
        S_{\rho\rho}(s,s',i\omega_n) &= S_{\rho\rho}^{(0)} (s,s',i\omega_n) \nonumber \\
        & \qquad +\beta\Phi(i\omega_n)A(s,s',i\omega_n) ,
\end{align}
where one has
\begin{align}
    \label{eq:S no QPC}
    S_{\rho\rho}^{(0)}(s,s',i\omega_n) 
    &= \frac{\nu}{2\pi L}\sum_{k\ne 0}\frac{k}{vk-i\omega_n}e^{ik(s-s')} ,\\
    \label{eq:Phi(i omega)}
    \Phi(i\omega_n) &= \frac{2}{\beta
    \Lambda_n}\left(-1 + \frac{\langle \phi_n\phi_{-n}\rangle}{2
    \beta
    \Lambda_n}\right) , \\
    \label{eq:A in charge dens}
    A(s,s',i\omega_n)&= \frac{\nu^2}{4L^2}\!\sum_{i,j=1}^{2} (-1)^{i+j} \nonumber \\
    & \qquad \times {\sum_{k,k'\neq 0}}\frac{e^{ik(s-s_i)}}{vk-i\omega_n}\frac{e^{-ik'(s'-s_j)}}{vk'-i\omega_n} .
\end{align}
In Eq.~\eqref{eq:Phi(i omega)}, we introduced the correlator for the local fields
\begin{equation}
\langle \phi_n\phi_{-n}\rangle = \frac{1}{Z[0]}\int \mathcal{D}\phi~\phi_n\phi_{-n} e^{-S_0[\phi]-S_{\rm tun}[\phi]}.
\end{equation}
For details of the calculation, see Appendix~\ref{app:density_correlator}.
Using the results of Eqs.~(\ref{eq:charge dens corr})–(\ref{eq:A in charge dens}), the integration in Eq.~(\ref{eq:permittivity}) yields the following response function in the Matsubara formalism
\begin{subequations}
\begin{align}
    \label{eq:whole epsilon}
    \epsilon(i\omega_n) ={}& \epsilon_0(i\omega_n) + \delta \epsilon(i\omega_n) \,,  \\
    \label{eq:epsilon 0}
    \epsilon_0(i\omega_n) ={}& \frac{\nu L^2}{16\pi^2}\left[\frac{1}{\omega_T + i\omega_n}+\frac{1}{\omega_T - i\omega_n}\right] , \\
    \label{eq:delta epsilon}
    \delta \epsilon(i\omega_n)
    ={}&  \frac{\beta \nu^2 L^2 }{16\pi^2}\Phi(i\omega_n)\Biggl[\frac{1}{(\omega_T-i\omega_n)^2} + \frac{1}{(\omega_T+i\omega_n)^2} \notag \\
    &- \frac{\cos{\frac{2\pi(s_1+s_2)}{L}}}{\omega_T}\!\left(\frac{1}{\omega_T + i\omega_n} + \frac{1}{\omega_T - i\omega_n}\right)\Biggr] .
\end{align}
\end{subequations}
Here $\epsilon_0(i\omega_n)$ corresponds to the response function of the free droplet without a QPC, while $\delta\epsilon(i\omega_n)$ characterizes the modification of the response function due to the quasiparticle tunneling at a QPC. In the absence of a QPC, $\delta\epsilon(i\omega_n)$ vanishes, and we see from the expression in Eq.~\eqref{eq:epsilon 0} that the resonance corresponds to a peak at frequency $\omega=\omega_T\equiv 2\pi v/L$,
which is the characteristic frequency of the droplet. We notice that, to obtain Eq.~\eqref{eq:delta epsilon}, we made the assumption $|s_1-s_2|=L/2$.

To obtain the response function in the
presence of tunneling at the QPC, one needs
to evaluate the bosonic correlator $\Phi(i\omega_n)$
[Eq.~\eqref{eq:Phi(i omega)}]. This cannot be done
exactly in the general case.
In this work, we use two different methods to 
obtain $\Phi(i\omega_n)$ [and thus $\epsilon(i \omega_n)$]: On the
one hand, we analytically calculate second-order perturbation in the tunneling strength $V$; on the
other hand, we use the numerical method of PIMC for arbitrary tunneling amplitudes. The PIMC method allows us not only to obtain results beyond the perturbative limit but also to confirm the perturbative results in their regime of validity.

From this point onward, and unless mentioned otherwise, we focus on a symmetric circular droplet, with a QPC connecting the positions $s_1=0$ and $s_2=L/2$ (see Fig.~\ref{fig:circular_model}).

\subsection{Perturbation theory}
\label{subsec:perturbation}

When the tunneling strength $V$ is small enough,
it is possible to perform a perturbative expansion
of the bosonic correlator $\Phi(i \omega_n)$ of Eq.~\eqref{eq:Phi(i omega)}. This leads to
\begin{align}
\epsilon(i \omega_n) &= \epsilon_0 (i \omega_n) +
 V \epsilon^{(1)}(i \omega_n)  + V^2 \epsilon^{(2)}(i \omega_n) + O(V^3)
 \label{eq:epsilon_perturb}
\end{align}
Since the zeroth-order response function $\epsilon_0$ exhibits narrow resonance peaks at $\pm\omega_T$, quasiparticle tunneling at the QPC is expected to change both the resonance frequencies and the peak widths.
However, a bare perturbative expansion of the form in Eq.~\eqref{eq:epsilon_perturb} does not allow one to directly extract these quantities, since the peak positions and linewidths are determined by the zeros of the inverse response function rather than by additive corrections to $\epsilon(i\omega_n)$ itself. 
To properly characterize these effects, we therefore introduce the interaction kernel $\Gamma(i\omega_n)$ as follows
\begin{align}
    \label{eq:Gamma def}
    \Gamma(i\omega_n)\equiv \epsilon^{-1}_0(i\omega_n) - \epsilon^{-1}(i\omega_n) .
\end{align}
Within this formulation, the real and imaginary parts of $\Gamma(\omega)$ directly encode the tunneling-induced peak shifts and the corresponding changes in the linewidth.
This represents a significant methodological distinction from the earlier analysis of Ref.~\cite{Cano2013PRB}, in which microwave absorption spectroscopy was proposed as a probe of quasiparticle tunneling and the tunneling effect was treated through a direct first-order perturbative expansion of the response function.
Within that treatment, the resonance-peak shifts were not discussed explicitly.
By contrast, our perturbative treatment of the interaction kernel $\Gamma(\omega)$ makes it possible to capture the peak shift due to quasiparticle tunneling.
Another important difference is that, in the present work, we include contributions up to second order in perturbation theory.
While the peak shift already appears at first order within our interaction-kernel formulation, changes in the linewidth emerge only at second order.
Considering these second-order contributions allows us to describe features intrinsic to fractional quantum Hall edge dynamics, such as the renormalization of the linewidth.

Using Eqs.~(\ref{eq:Phi(i omega)}), (\ref{eq:delta epsilon}), and (\ref{eq:epsilon_perturb}), we obtain
for $\Gamma(i \omega_n)$ up to second order in $V$
\begin{widetext}
\begin{align}\label{eq:Gamma(2)}
        \Gamma^{(2)}(i\omega_n)= -V\frac{64\pi^2}{L^2}e^{-\bar{\Lambda}}
        - V^2\frac{32\pi^2e^{-2\bar{\Lambda}}}{L^2}\biggl[8\nu\left(\frac{1}{\omega_T-i\omega_n} + \frac{1}{\omega_T+i\omega_n}\right)+ \mathcal{P}_{+}(0) + \mathcal{P}_{-}(0) - \mathcal{P}_{+}(i\omega_n) + \mathcal{P_{-}}(i\omega_n) - 2\beta\biggr].
\end{align}
\end{widetext}
Here, $\mathcal{P}_{\pm}(i\omega_n)$ and $\bar{\Lambda}$ are defined as 
\begin{align}
    \label{eq:P_pm def}
    \mathcal{P}_{\pm}(i\omega_n) &\equiv \int_0^{\beta}d\tau\,e^{i\omega_n\tau}e^{\pm\frac{2}{\beta}\sum_{p}\Lambda_p\cos{\omega_p\tau}}, \\
    \bar{\Lambda} &\equiv \frac1\beta\sum_{\omega_m}\Lambda_m,
\end{align}
and one can show that $e^{-\bar{\Lambda}} \simeq 
(\pi \alpha/L)^{\nu}$.
Once $\Gamma^{(2)}$ is computed, the response function $\epsilon_p (i\omega_n)$ in the perturbative regime is then obtained as
\begin{equation}
    \epsilon_{\rm p}(i\omega_n) = \frac{1}{\epsilon_0^{-1}(i\omega_n) - \Gamma^{(2)}(i\omega_n)}.\label{eq:epsilon up to 2nd perturb}
\end{equation}
After performing the analytic continuation to the real frequency, the real part of $\Gamma^{(2)} (\omega + i \delta)$ corresponds to the shift in the position
of the resonance peaks, while the imaginary
part yields the broadening and deformation of the peaks.
From Eq.~\eqref{eq:Gamma(2)}, one readily sees that the first order term in $\Gamma^{(2)} (\omega+i\delta)$ is purely real, thus leading to a simple shift of the resonance peaks,
while the second order term yields both a shift and a widening. Note that the order $n$ term contains a factor
$e^{-n \bar{\Lambda}} = (\pi \alpha/L)^{\nu n}$.
As $ (\pi \alpha/L) \ll 1$, this is a small factor that guarantees the validity of the perturbation expansion. As this factor increases when $\nu$
decreases, one can expect that higher-order terms in the perturbative expansion will be more important
for a given $V$ as $\nu$ decreases (see Sec.~\ref{sec:FQH result}).

\subsection{PIMC method}
\label{subsec:PIMC}

In this study, we also evaluate $\Phi(i\omega_n)$ using the PIMC method and thereby obtain a numerically exact response function.
In this approach, the partition function is represented as an imaginary-time path integral over $\phi(\tau)$, and representative paths are sampled with the weight determined by the Euclidean action.
From the $\Phi(i\omega_n)$ obtained via PIMC, we compute $\Gamma(i\omega_n)$ and perform numerical analytic continuation to finite real frequencies using the Pad\'{e} approximation, thereby obtaining the response function on the real-frequency axis from Eq.~\eqref{eq:Gamma def}.
Since the PIMC results include higher-order effects in the coupling strength, this method enables us to access the response not only in the region dominated by second-order perturbations but also in regimes where higher-order effects become significant

The PIMC procedure is summarized as follows. In the numerical implementation, the imaginary-time interval $[0,\beta)$ is discretized as $\tau_l=l\Delta\tau$ ($l=0,1,2,\dots,N-1$), with $\Delta\tau=\beta/N$, and the path is represented by $\phi(\tau_l)$.
With this discretization, the effective action is approximated as
\begin{align}
    S_{\rm eff}[\phi] &= S_{0}[\phi] + S_{\rm tun}[\phi], \\   
    S_{0}[\phi] &= \sum_{n=-N/2+1}^{N/2} \frac{|\phi_n|^2}{4\beta \Lambda_n}, \\
    \label{eq:discretized S_tun}
    S_{\rm{tun}}[\phi] &= -V \Delta\tau \sum_{l=0}^{N-1} \cos\left[\phi(\tau_l)\right] .
\end{align}
The Fourier transform defined in Eq.~(\ref{eq:phiFourier}) is accordingly modified to
\begin{align}
    \label{eq:discretized phiFourier}
    \phi(\tau_l) ={}& \frac{1}{\beta}\sum_{n=-N/2+1}^{N/2} \phi_n e^{-i\omega_n\tau_l}.
\end{align}
Here, the Fourier components satisfy the relation $\phi_{-n} = \phi_n^*$.
The Monte Carlo update is carried out in the frequency domain. Starting from a configuration $\{\phi_n\}$, we successively choose a frequency index $k$ in the range $[0,\,N/2]$ and propose a new value of $\phi_k$ drawn from the Gaussian distribution associated with the quadratic action $e^{-S_0[\phi]}$.
The original and proposed configurations are then transformed to the imaginary-time domain, and the change in the tunneling action is evaluated as
\begin{align}
\Delta S = S_{\rm{tun}}[\{ \phi(\tau_l) \}^{\rm (new)} ] - S_{\rm{tun}}[\{ \phi(\tau_l) \}] .
\end{align}
The proposed update is accepted according to the Metropolis criterion with probability $\min\left(1,e^{-\Delta S}\right)$. A complete sweep over all frequency components constitutes one Monte Carlo step. Repeating this procedure generates imaginary-time paths distributed according to the Boltzmann weight $e^{-S_{\rm eff}[\phi]}$, from which the correlation functions are evaluated by importance sampling.

\section{Results for the IQH droplet case}
\label{sec:IQHE}

In this section, we present the analytical calculation for the IQH droplet with a filling factor $\nu=1$, where the edge state is described by a noninteracting chiral electron system. 
We benchmark the consistency of the perturbative and PIMC results against the analytic solution.

\begin{figure}[tb]
\begin{center}
\includegraphics[clip,width=7.0cm]{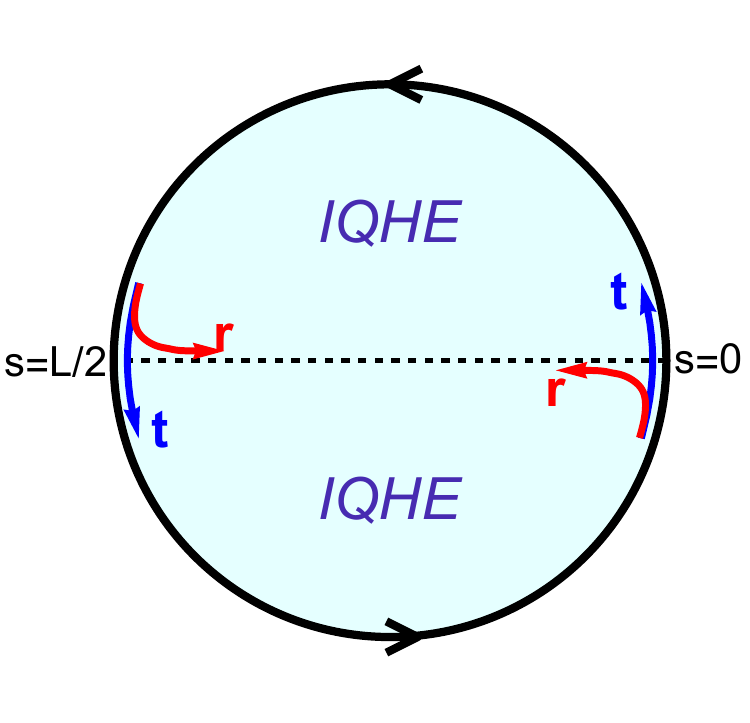}
\caption{Schematic and notation used in the exact IQHE calculation. 
The droplet is circular, with perimeter $L$. 
The QPC connects the points at $s=0$ and $s=L/2$, and is characterized by reflection and transmission amplitudes $r$ and $t$.}
\label{fig:setupScat}
\end{center}
\end{figure}

We consider a finite edge state in an IQH droplet, as illustrated in Fig.~\ref{fig:setupScat}.
At the QPC, which connects the points $s=0$ and $s=L/2$, electrons can either be transmitted or reflected to the opposite point. 
We define the $S$ matrix at the QPC as
\begin{align}
    \hat{S} = \left(
    \begin{matrix}
        t & -ir \\
        -ir & t
    \end{matrix}
    \right) ,
    \label{eq:Smatrix}
\end{align}
where $r$ and $t=\sqrt{1-r^2}$ are the reflection and transmission amplitudes, respectively (see Fig.~\ref{fig:setupScat}), which are determined from the quasiparticle tunneling $\lambda_r$ (see Appendix~\ref{app:IQHE}).
We note that $t=1$ corresponds to the absence of a QPC ($\lambda_r = 0$).
As mentioned in Sec.~\ref{subsec:formulation eff action}, a uniform magnetic field is applied to the droplet, and as electrons move along the edge, they acquire a phase factor $e^{i\phi_{\rm AB}s/L}$ due to the AB effect. The AB phase is given by $\phi_{\rm AB} = 2\pi \Phi / \Phi_0$, as stated previously, where the flux quantum is $\Phi_0 = h/e$. 

By imposing the twisted boundary conditions that account for the AB effect, the wavenumber is discretized as
\begin{subequations}
\begin{align}
    \label{eq:k shift2}
    k_{n,j} &= \frac{4\pi}{L}(n+\delta_j), \quad (n\in \mathbb{Z},\ j=1,2),\\
    \delta_1 &= \frac{\chi}{2\pi} - \frac{\phi_{\rm AB}}{4\pi}, \\
    \delta_2 &= \frac{1}{2}-\frac{\chi}{2\pi} - \frac{\phi_{\rm AB}}{4\pi},
\end{align}
\end{subequations}
where $\chi = \arctan(r/t) \in [0,\pi/2]$ denotes the coupling strength at the QPC (for a detailed derivation, see Appendix~\ref{app:IQHE}).
In the absence of the QPC ($r=0$), the energy-level spacing is simply given by an integer multiple of $2 \pi/L$, reflecting the edge-state length $L$. 
When the QPC is introduced, the energy levels are divided into two groups $k_{n,1}$ and $k_{n,2}$, having their wavenumbers shifted in opposite directions by a factor $\chi/2 \pi$.
The AB phase $\phi_{\rm AB}$ induces uniform shifts of both wavenumbers $k_{n,1}$ and $k_{n,2}$, which have the same effect as changing the chemical potential.

The electrons that pass through the QPC at $s = L/2$ acquire $\Theta_j=(-1)^{j}\chi$ for the wavenumber $k_{n,j}$ (see Appendix~\ref{app:IQHE}).
Accordingly, the imaginary-time evolution of the electron field operator $\psi(s,\tau)$ is expressed as
\begin{align}
    \label{eq:ele ope}
    \psi(s,\tau) &= \frac{1}{\sqrt{L}}\sum_{n}\sum_{j=1,2} e^{ik_{n,j}s-vk_{n,j}\tau} g_j(s) c_{n,j}, \\
    g_j(s) &= \left\{ \begin{array}{cc} 1, & (s<L/2) ,\\ e^{i\Theta_j}, & (s>L/2) , \end{array} \right.  .
\end{align}
The imaginary-time response function is given by
\begin{align}
    \label{eq:ele dens corr}
    S_{\rho\rho} (s,s',\tau) = \langle\psi^{\dagger}(s,\tau)\psi(s',0)\rangle\langle\psi(s,\tau)\psi^{\dagger}(s',0)\rangle.
\end{align}
for $0<\tau<\beta$.
By substituting Eqs.~\eqref{eq:ele ope}-\eqref{eq:ele dens corr} into Eq.~\eqref{eq:permittivity} and performing the integration over $s$ and $s'$, the imaginary part of the susceptibility is calculated as
\begin{align}
    \label{eq:ele epsilon exact}
    &{\rm Im}\, \epsilon^R(\omega) 
    =\frac{L^2 \sin^2 \chi}{16 \pi^3} \notag \\
    &\times \sum_{m,n}
    \frac{f(vk_{n,1})-f(vk_{m,2})}{(n-m+\chi/\pi)^2(n-m+\chi/\pi-1)^2} \notag \\
    &\times \Bigl\{\delta\qty[\omega +v(k_{n,1}-k_{m,2})] - \delta\qty[\omega -v(k_{n,1}-k_{m,2})]\Bigr\}  .
\end{align}
The response function is determined by transitions between the different eigenstates with eigenenergies $vk_{n,1}$ and $vk_{m,2}$, as indicated by the delta function $\delta\qty[\omega \pm v(k_{n,1}-k_{m,2})]$ (see also Appendix~\ref{app:IQHE}). 
Due to the factor $f(vk_{n,1})-f(vk_{m,2})$, only transition from occupied states to unoccupied states are possible at zero temperature.

\begin{figure}[tb]
\begin{center}
\includegraphics[clip,width=9.2cm]{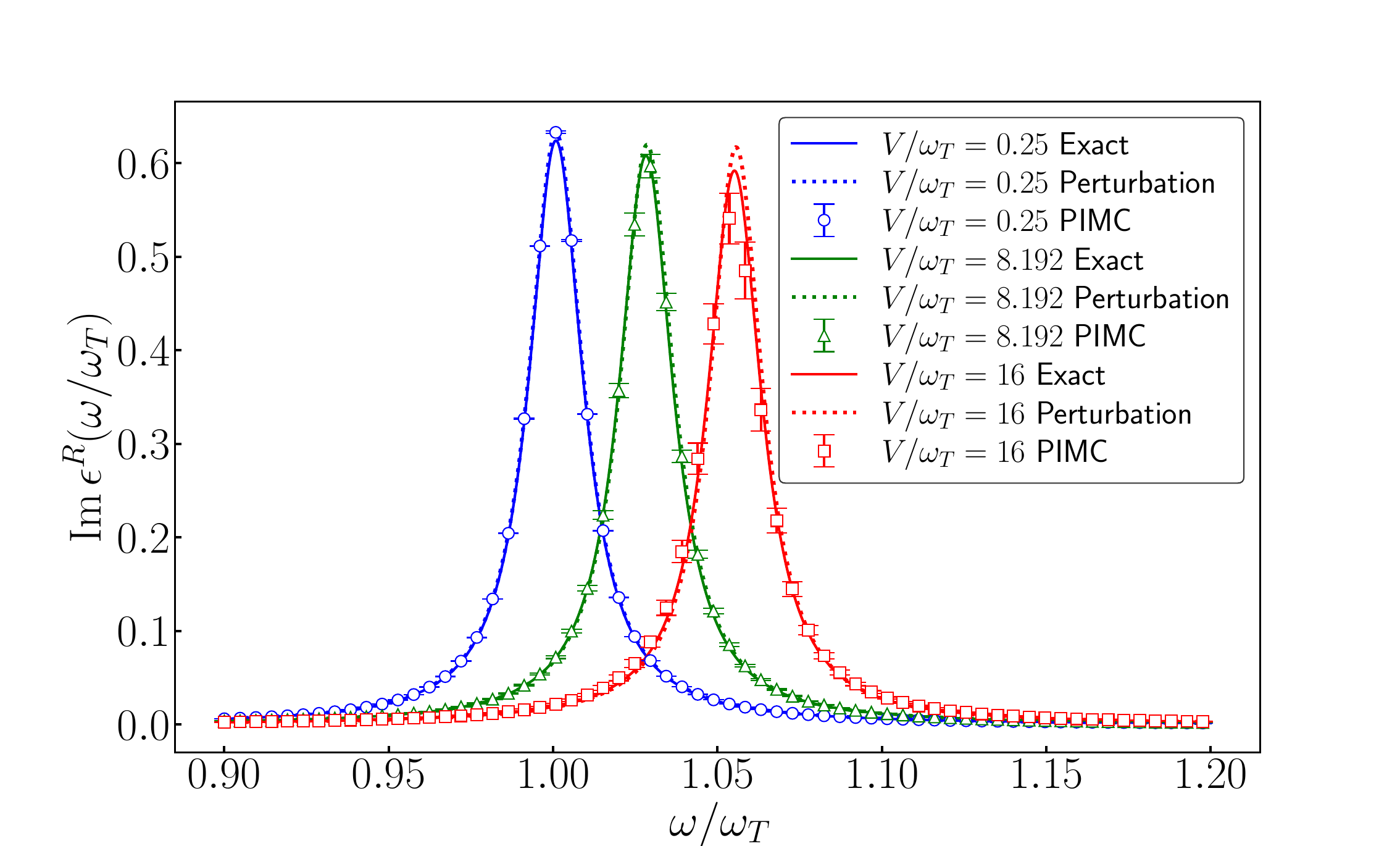}
\caption{Imaginary part of the susceptibility ${\rm Im}\,\epsilon^{R}(\omega)$ for the $\nu=1$ IQH droplet, shown in units of $L^2/\omega_T$. 
The temperature is set to $T/\omega_T = 0.1$. 
Solid and dashed lines show the results of scattering theory [Eq.~\eqref{eq:ele epsilon exact}] and second-order perturbation theory [Eqs.~\eqref{eq:Gamma(2)}-\eqref{eq:epsilon up to 2nd perturb}], respectively, while symbols with error bars represent numerical data obtained using the PIMC method. 
Dimensionless coupling strengths are $V/\omega_T = 0.25$, $8.192$, and $16$. 
For visual clarity, delta functions are broadened into Lorentzians with linewidth $\delta = \omega_T/100$.}
\label{fig:nu=1_comparison}
\end{center}
\end{figure}

This exact expression allows us to check the correctness of our two calculation methods: the second-order perturbation and the PIMC simulation.
To make comparison easier, the delta function in Eq.~\eqref{eq:ele epsilon exact} is replaced with a Lorentzian function with a small width $\delta$.
This corresponds to the analytic continuation $i\omega_n \rightarrow \omega + i\delta$ with a finite value of $\delta$.
Figure~\ref{fig:nu=1_comparison} shows the response function ${\rm Im}\,\epsilon^{R}(\omega)$ for the case
of a circular droplet at $\nu=1$. It compares the results of all three methods: the exact theoretical treatment of Eq.~\eqref{eq:ele epsilon exact}, the results obtained from second-order perturbation theory [Eqs.~\eqref{eq:Gamma(2)}-\eqref{eq:epsilon up to 2nd perturb}], and the numerical results computed using the PIMC method.
In this comparison, a clear shift of the peak at $\omega = \omega_T = 2\pi v/L$ toward higher frequencies is observed as the coupling strength increases, and all three results show good overall agreement within the range of coupling strengths considered.
For larger coupling strengths, particularly at $V/\omega_T = 16$, the Monte Carlo and exact theoretical results remain in good agreement, whereas the second-order perturbation begins to show a slight deviation in intensity.
We found that second-order perturbation agrees well with the exact theoretical treatment for moderate coupling strengths, while the PIMC simulations accurately reproduce the exact theoretical results over the entire range considered.

\section{Results for the FQH droplet Case}
\label{sec:FQH result}

Next, we discuss the behavior of ${\rm Im}\,\epsilon^{R}(\omega)$ for a FQH droplet, based on perturbation theory and numerical calculations using the PIMC method, as an exact analytical calculation is not available in this case. 
Through this comparison, we further examine how the deviation between the perturbative and PIMC results depends on the filling factor.

\begin{figure}[tb]
\begin{center}
\includegraphics[clip,width=9.2cm]{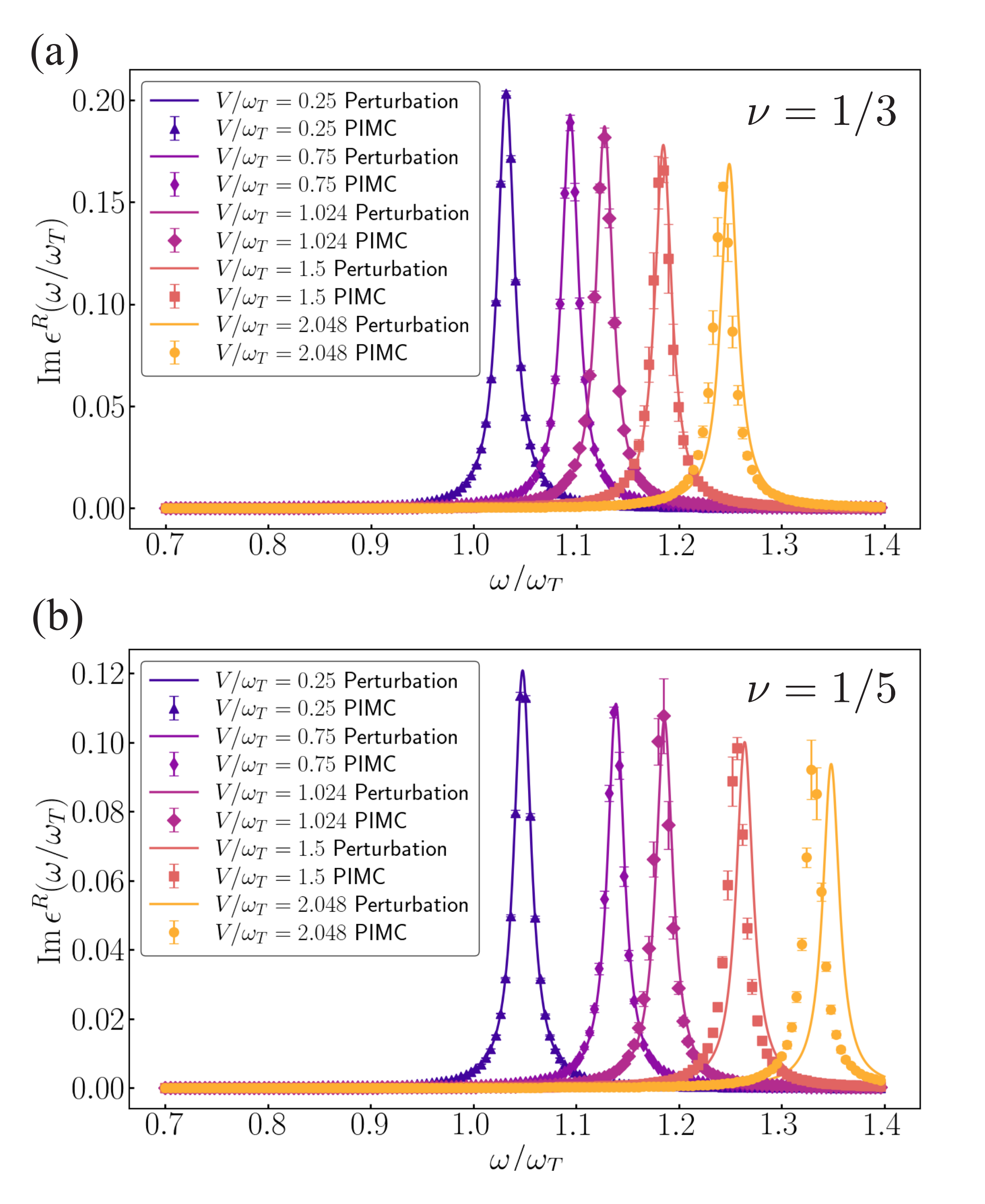}
\caption{Imaginary part of the susceptibility ${\rm Im}\,\epsilon^{R}(\omega)$ for FQH droplets with (a) $\nu=1/3$ and (b) $\nu=1/5$ at temperature $T/\omega_T = 0.1$. 
Solid lines show results from second-order perturbation theory, whereas symbols with error bars represent data obtained using PIMC. 
Coupling strengths are $V/\omega_T = 0.25, 0.75, 1.024, 1.5,$ and $2.048$. 
A finite broadening $\delta = \omega_T/100$ is used.}
\label{fig:FQH_vdep}
\end{center}
\end{figure}

Figures~\ref{fig:FQH_vdep}(a) and \ref{fig:FQH_vdep}(b) show the behavior of ${\rm \Im}\,\epsilon^{R}(\omega)$ for $\nu=1/3$ and $1/5$, respectively, with varying dimensionless coupling strength $V/\omega_T = 0.25$, $0.75$, $1.024$, $1.5$, and $2.048$ at a fixed temperature of $T/\omega_T = 0.1$. 
To make the peak shift clearer, we have used a finite broadening factor $\delta = \omega_T/100$ in the analytic continuation.
The solid lines represent the perturbative results, while the plotted points denote the PIMC results. 
As in the case of the IQH droplet, the resonance peak shifts toward higher frequencies as the coupling strength increases. In both figures, this peak shift is more pronounced than in the IQH droplet.
Moreover, for a given value of the coupling strength $V$, the shift is found to be larger for $\nu=1/5$ than for $\nu=1/3$. As seen from Eq.~\eqref{eq:Gamma(2)}, the first- and second-order terms are accompanied by factors of $e^{-\bar{\Lambda}}$ and $e^{-2\bar{\Lambda}}$, respectively.
These factors get larger as $\nu$ decreases since $e^{-\bar{\Lambda}} \simeq \left( \pi \alpha/L\right)^\nu$.
Therefore, it can be concluded that the effect of the coupling strength becomes more significant for smaller $\nu$.

\begin{figure}[tb]
\begin{center}
\includegraphics[clip,width=9.2cm]{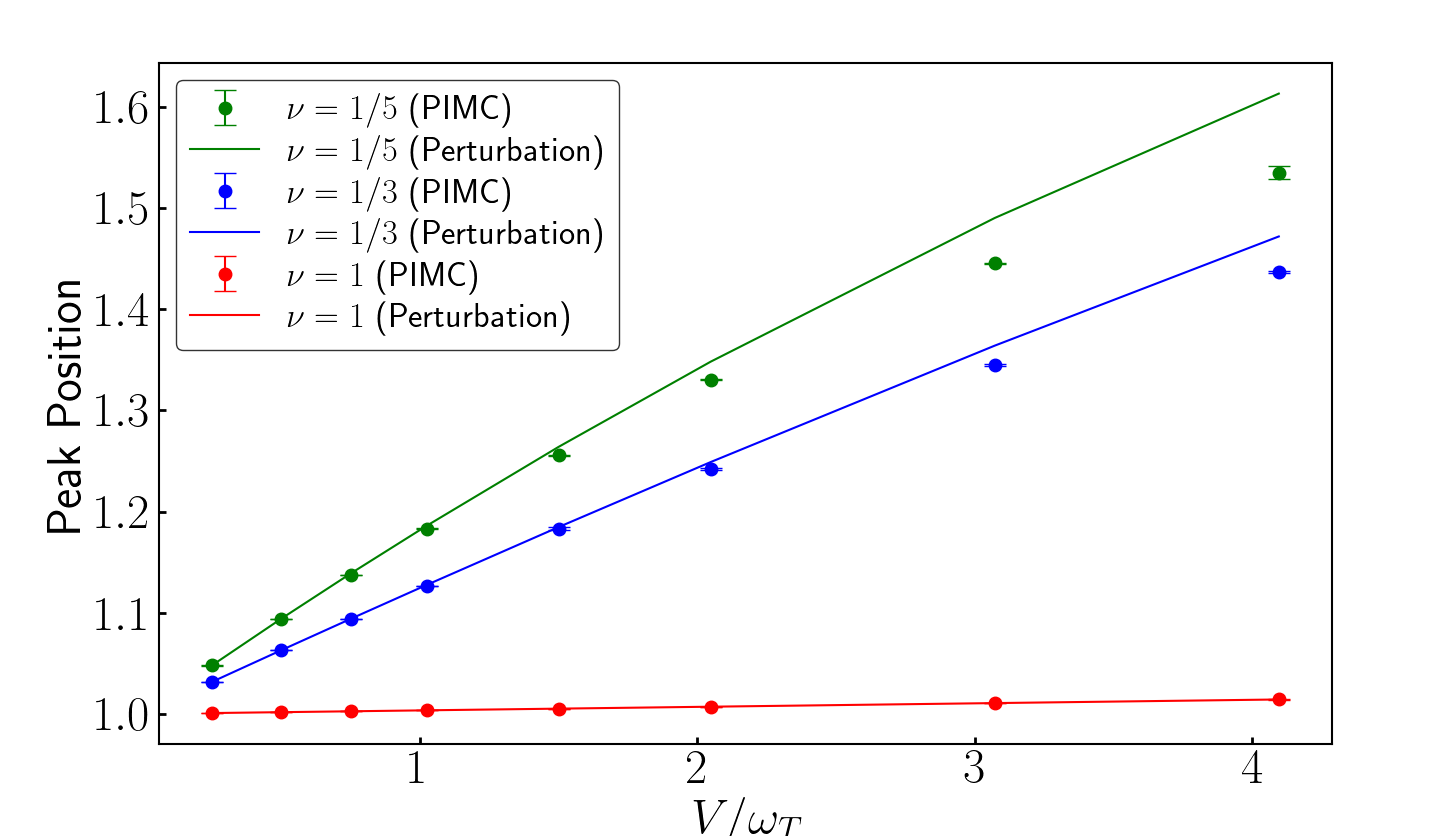}
\caption{Dependence of resonance-peak positions on tunneling strength for filling fractions $\nu = 1/5$, $1/3$, and $1$. 
Solid curves denote results from second-order perturbation theory, while symbols represent PIMC results.}
\label{fig:peak_position}
\end{center}
\end{figure}

In Fig.~\ref{fig:FQH_vdep}, as the coupling strength increases, a noticeable deviation between the PIMC and second-order perturbative results begins to appear. This deviation emerges at smaller coupling strengths for $\nu = 1/5$ than for $\nu = 1/3$, indicating that the $\nu = 1/5$ state is more sensitive to the coupling strength and exhibits higher-order effects more prominently. Figure~\ref{fig:peak_position} shows the dependence of the peak position on the coupling strength for $\nu = 1$, $1/3$, and $1/5$.
The solid lines represent the peak shifts obtained from the second-order perturbation results, while the plotted points correspond to those obtained from the PIMC calculations.
Within the range of coupling strengths shown, the peak for $\nu = 1$ exhibits only a small shift, and the perturbative and PIMC results agree well.
As discussed above, the peak shift becomes larger as $\nu$ decreases from $1/3$ to $1/5$.
As the coupling strength increases, a noticeable deviation appears between the perturbative and PIMC results, with the deviation being more significant for $\nu = 1/5$ than for $\nu = 1/3$.
This observation suggests that higher-order effects become increasingly important for smaller filling factors and that such effects are successfully captured by the PIMC calculations.

\begin{figure}[tb]
\centering
\label{fig:nu=1/3_Tdep}
\includegraphics[clip,width=9.2cm]{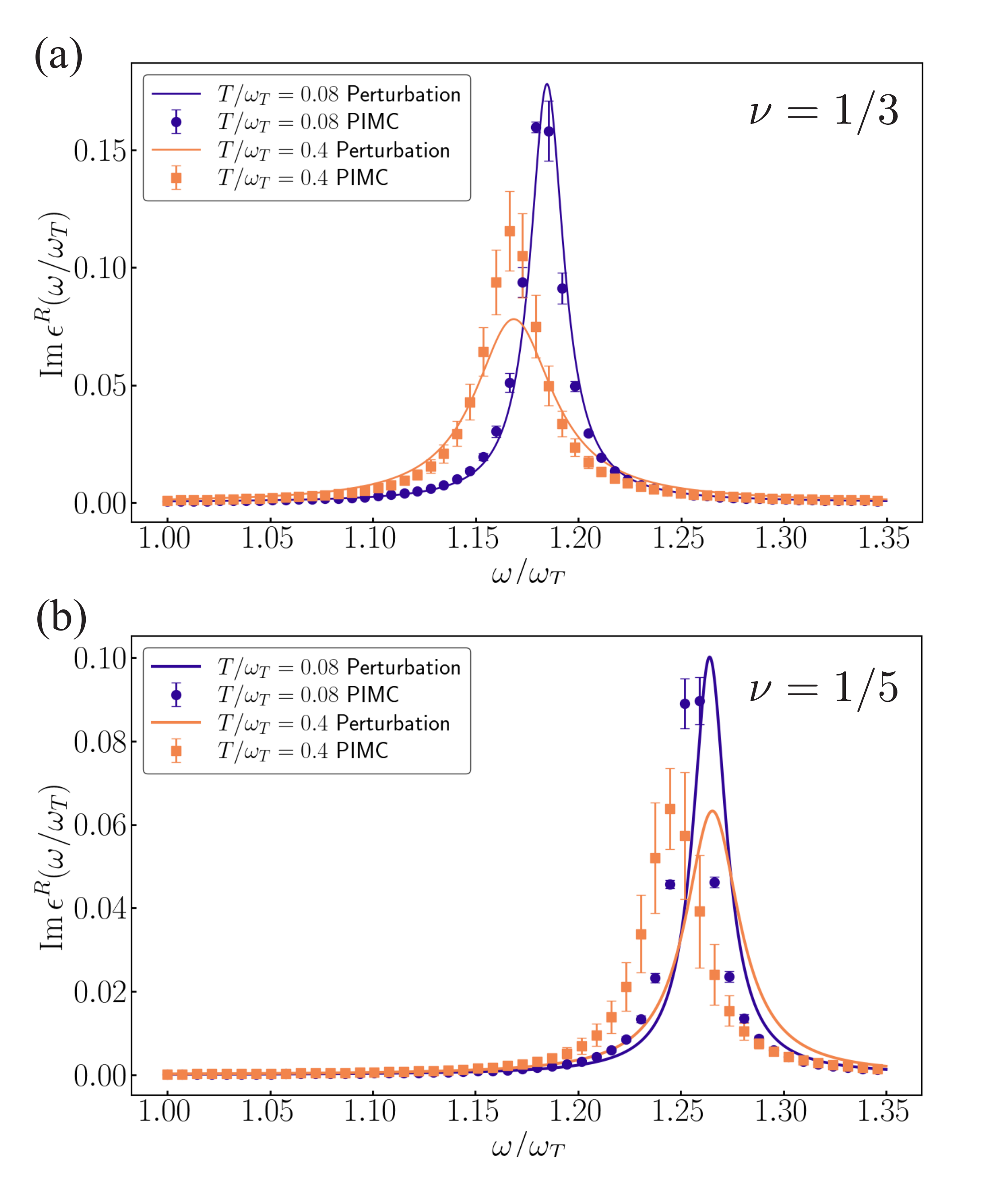} \caption{Imaginary part of the susceptibility ${\rm Im}\,\epsilon(\omega)$ at $T/\omega_T = 0.08$ (blue) and $0.4$ (orange) for FQH droplets with (a) $\nu = 1/3$ and (b) $\nu = 1/5$ at coupling strength $V/\omega_T = 1.5$. 
A broadening factor $\delta = \omega_T/100$ is used.}
\label{fig:FQH_Tdep}
\end{figure}

Next, Figs.~\ref{fig:FQH_Tdep}(a) and \ref{fig:FQH_Tdep}(b) show the temperature dependence of ${\rm Im}\,\epsilon^{R}(\omega)$ for $\nu=1/3$ and $1/5$. The coupling strength is fixed at $V/\omega_T=1.5$. As in Fig~\ref{fig:FQH_vdep}, the solid lines represent the perturbative results, while the plotted points denote the PIMC results.
For both $\nu = 1/3$ and $1/5$, the increase in linewidth with rising temperature is commonly observed in both the perturbative and PIMC results. In the present formulation, phonon effects are not considered, indicating that the linewidth broadening is caused by thermal fluctuations. In the PIMC results, a low-energy shift with increasing temperature is observed for both $\nu = 1/3$ and $1/5$. While the perturbative result for $\nu = 1/3$ exhibits a similar behavior, the one for $\nu = 1/5$ shows no noticeable peak shift as the temperature increases. From this result, the low-energy peak shift observed for $\nu = 1/5$ is considered to be caused by higher-order effects.

\section{Results for the realistic geometry case}
\label{sec:RealisticGeometry}

\begin{figure}[tb]
\begin{center}
\includegraphics[clip,width=8cm]{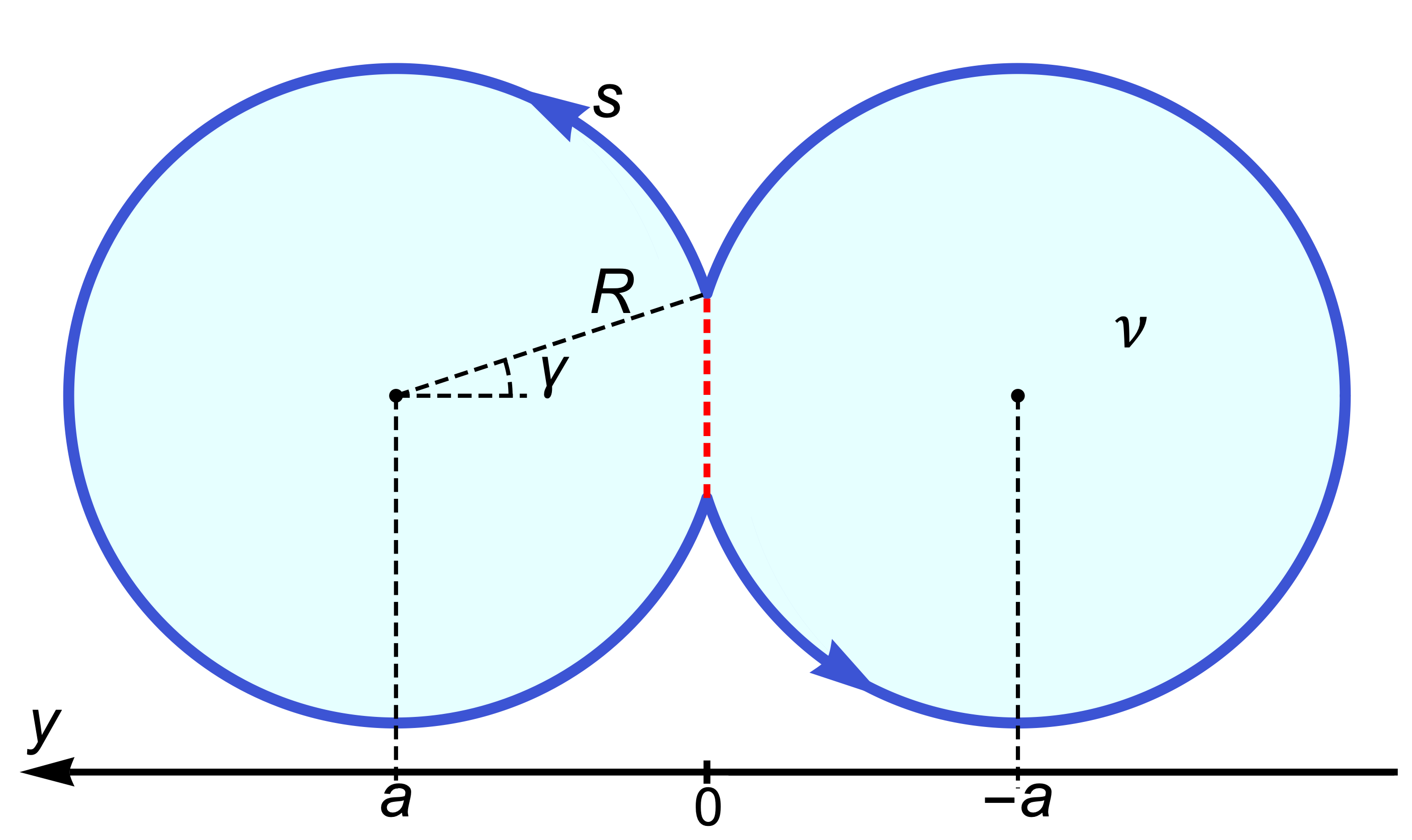}
\caption{Schematic of a realistic droplet geometry composed of two circular arcs joined at a junction. 
The QPC is formed by a constriction at this junction, whose opening angle is denoted by $\gamma$.}
\label{fig:real_geometry}
\end{center}
\end{figure}

We now present the dielectric response in a more realistic geometry. 
So far, for simplicity, we have focused on a circular droplet geometry. 
In the following, we consider a geometry constructed from two overlapping circles of equal radius $R$, whose centers are laterally offset by $2a$, as illustrated in Fig.~\ref{fig:real_geometry}. 
Equation~(\ref{eq:permittivity}) provides the general expression for the response function, and by using the edge profile $y(s)$ corresponding to this droplet shape, the calculation can be carried out in the same manner as in the circular case. 
Importantly, the geometry dependence enters only through the edge profile $y(s)$, while the bosonic correlation function is geometry independent. 
Therefore, responses for different geometries can be obtained using a single set of Monte Carlo data for the bosonic correlator, without repeating the numerically expensive simulations.

In Fig.~\ref{fig:real_geometry}, the radius of each circle $R$ is given by $R = L/[4(\pi - \gamma)]$, as obtained from the geometry shown in the figure. Furthermore, from the relation $R\cos{\gamma} = a$, it follows that the angle $\gamma$ is determined once the distance between the centers of the two circles is specified. At this point, $y(s)$ can be expressed as follows:
\begin{align}
    \label{eq:real_y}
    y(s) &= \left\{ \begin{array}{cc} R\left[\cos{\gamma} -\cos\qty(R^{-1}s + \gamma)\right], & (s<L/2) ,\\ -R\left[\cos{\gamma} -\cos\qty(R^{-1}s + 3\gamma)\right], & (s>L/2) , \end{array} \right.  .
\end{align}
By performing the integration in Eq.~(\ref{eq:permittivity}), the response function in Matsubara space is obtained as follows:
\begin{subequations}
\begin{align}
    \label{eq:whole epsilon _real}
    \epsilon^{\rm{real}}(i\omega_n) &= \epsilon^{\rm{real}}_0(i\omega_n) + \delta \epsilon^{\rm{real}}(i\omega_n) \,,  \\
    \label{eq:epsilon_real 0}
    \epsilon^{\rm{real}}_0(i\omega_n) &= \frac{8\nu  R^{-2}\cos^2{\gamma}}{\pi L}\notag \\
    &\times \sum_{l\in\mathbb{Z}}\frac{1}{\Delta_l\qty(R^{-2}-\Delta_l^2)^2\qty(v\Delta_l-i\omega_n)}, \\
    \label{eq:delta epsilon_real}
    \delta \epsilon^{\rm{real}}(i\omega_n)
    &= \frac{16\beta\nu^2 R^{-2} \cos^2{\gamma}}{L^2}\Phi(i\omega_n)\notag \\ &\times\Biggl[\sum_{l\in\mathbb{Z}}\frac{1}{\Delta_l\qty(R^{-2}-\Delta_l^2)\qty(v\Delta_l-i\omega_n)}\Biggr]^2 ,
\end{align}
\end{subequations}
where $\Delta_l\equiv2(2l-1)\pi/L$. The resulting susceptibility displays a set of peaks at energies set by $v\Delta_l$, with $l\in\mathbb{Z}$. Therefore, since $v\Delta_l = (2l - 1)\omega_T$, it follows that in the geometry shown in Fig.~\ref{fig:real_geometry}, multiple peaks appear at odd multiples of $\omega_T$. This geometry reduces to that of the circular droplet with QPCs located at $s = 0$ and $s = L/2$ when $\gamma = \pi/2$.
Indeed, Eqs.~(\ref{eq:epsilon_real 0}) and (\ref{eq:delta epsilon_real}) coincide with Eqs.~(\ref{eq:epsilon 0}) and (\ref{eq:delta epsilon}) in the limit $\gamma \rightarrow \pi/2$ for $s_1+s_2=L/2$.

\begin{figure}[tb]
\begin{center}
\includegraphics[clip,width=9.2cm]{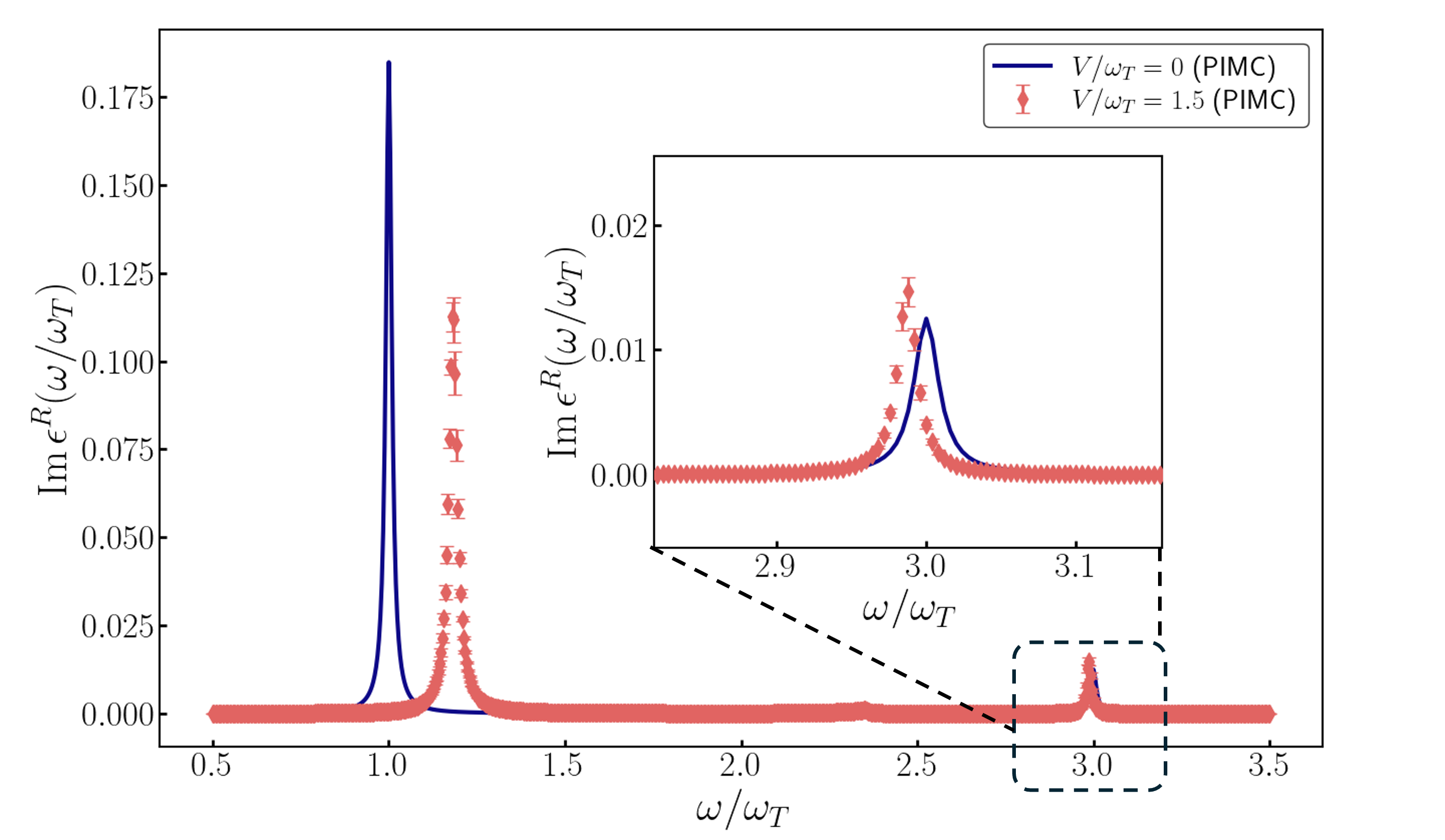}
\caption{Imaginary part of the susceptibility ${\rm Im}\,\epsilon^{R}(\omega)$ at $T/\omega_T = 0.1$ for a $\nu = 1/3$ FQH droplet with the geometry shown in Fig.~\ref{fig:real_geometry}. 
The solid curve shows the analytical result at zero tunneling strength $V/\omega_T = 0$, while symbols represent PIMC data at $V/\omega_T = 1.5$. 
The constriction of the droplet is controlled by the geometric angle $\gamma$, which is set to $\gamma = \pi/15$. 
A broadening $\delta = \omega_T/100$ is used.
The inset shows a magnified view of the secondary peak.}
\label{fig:real_geometry_peak_nu13}
\end{center}
\end{figure}

Figure~\ref{fig:real_geometry_peak_nu13} shows the calculated $\rm{\Im}\,\epsilon^R(\omega)$ for the $\nu = 1/3$ FQH droplet with the geometry illustrated in Fig.~\ref{fig:real_geometry}.
It is confirmed that the secondary peak appears at three times the frequency of $\omega_T$.
As indicated from Eqs.~(\ref{eq:epsilon_real 0}) and (\ref{eq:delta epsilon_real}), the amplitude decays rapidly when increasing the mode index $l$.
In addition, as the coupling strength increases, a peak shift occurs, similar to the case of the circular droplet. The direction of the shift depends on the peak position.
The peak located at $\omega/\omega_T = 1$ shifts toward higher frequencies, while the one at $\omega/\omega_T = 3$ shifts toward lower frequencies.

\section{Summary}
\label{sec:Summary}

In this work, we investigated the microwave response of FQH droplets containing a QPC using a combination of PIMC simulations and perturbation theory up to second order in the tunneling strength. 
For IQH droplets, we benchmarked the approach against results obtained from quantum scattering theory, confirming the consistency of the response functions in regimes where analytical solutions are available.

We demonstrated that quasiparticle tunneling induces resonance peak shifts as well as thermal linewidth broadening, which were not addressed in the original proposal of Ref.~\cite{Cano2013PRB}. 
We further showed that these peak shifts can be discussed reasonably only when the interaction kernel of the response function is considered, highlighting the limitations of bare perturbative expansions. 
The resulting behavior exhibits a clear dependence on the filling factor, reflecting the evolution of correlation strength in FQH states and explaining the increasing deviation between PIMC and finite-order perturbative results at smaller filling factors.

Because the PIMC framework naturally incorporates higher-order contributions beyond perturbation theory while retaining geometric and thermal flexibility, it provides a robust tool for evaluating microwave response in regimes where perturbative approaches become inaccurate.
These results establish microwave spectroscopy as a quantitative probe of quasiparticle dynamics and collective edge excitations in mesoscopic quantum Hall droplets, and offer a practical pathway for connecting theoretical modeling with experimental measurements across a wide range of filling factors and device geometries. The proposed spectroscopy protocol is compatible with current quantum Hall experimental platforms, particularly in view of recent developments in gate-tunable edge-magnetoplasmon resonators and the high degree of control achieved over QPCs~\cite{Frigerio2024CommPhys}.
Within the model considered here, quasiparticle (anyon) braiding effects are not the dominant mechanism influencing the microwave response. 
Extending the PIMC framework to regimes where braiding plays a central role in microwave resonances represents a promising direction for future investigation. Another natural extension of the present work is the investigation of FQH states with multiple edge channels, such as those of the Jain sequence, where interchannel interactions and charge fractionalization may lead to a richer microwave absorption spectrum.

\section*{Acknowledgements}

We thank G. F\`eve, G. M\'enard and F. Parmentier for useful discussions on experimental aspects. F.M. is supported by the International Graduate Program of Innovation for Intelligent World of the University of Tokyo. This research is supported by JST SPRING, Grant No.~JPMJSP2108. R.S. is supported by the Hirose
Foundation and the Murata Science and Education Foundation. T.K. acknowledges the support of the Japan Society for the Promotion of Science (JSPS KAKENHI Grant No.~JP24K06951). F.M. and T.K. acknowledge the support of the JST ASPIRE Program No.~JPMJAP2410.
This work was carried out in the framework of the
project ``ANY-HALL'' (Grant No. ANR-21-CE30-0064-03). We acknowledge
funding from the Agence Nationale de la Recherche under
the France 2030 programme, Reference No. ANR-22-PETQ-0012, and from the ERC Advanced Grant
``ASTEC'' (Grant No. 101096610). This French-Japanese collaboration is supported by the
CNRS International Research Project ``Excitations in Correlated Electron Systems driven in the GigaHertz range''
.

\section*{Data Availability Statement}

The data and code that support the findings of this article are openly available in Ref.~\cite{Murabayashi2026zenodo}.

\appendix

\section{Derivation of the effective action}
\label{app:effective_action}

In this appendix, we present the procedure for deriving the effective action.
We first introduce the Fourier transforms of the bosonic field $\varphi(s,\tau)$ and the auxiliary field $\lambda(\tau)$ as
\begin{align}
    \varphi(s,\tau) &=  \frac{1}{L\beta}\sum_{k\ne 0}\sum_{\omega_n}
    \varphi_{k,n} e^{iks-i\omega_n\tau} , \\
    \lambda(\tau) &= \frac{1}{\beta}\sum_{\omega_n}
    \lambda_n e^{-i\omega_n\tau} .
\end{align}
Using these definitions, the imaginary-time actions
$S_{\rm edge}[\varphi]$ and $S_{\lambda}[\varphi,\lambda,\phi]$,
originally defined in Eqs.~(\ref{eq:S_edge}) and (\ref{eq:S_lambda}),
respectively, can be rewritten as
\begin{align}
    S_{\rm edge}[\varphi] &=
    \frac{1}{L\beta}\sum_{k\ne 0} \sum_{\omega_n}
    \frac{k(vk - i\omega_n)}{4\pi \nu}
    \left|\varphi_{k,n}\right|^2, \\
    S_{\lambda}[\varphi,\lambda,\phi] &=
    \frac{i}{L\beta} \sum_{k\ne 0} \sum_{\omega_n}
    \lambda_{-n} \notag \\
    & \hspace{5mm} \times 
    \left[(e^{iks_1}-e^{iks_2})\varphi_{k,n} - L\phi_n \right] .
\end{align}
Since the total action $S_{\rm edge}[\varphi]
+ S_{\lambda}[\varphi,\lambda,\phi]
+ S_{\rm tun}[\phi]$ is quadratic in $\varphi_{k,n}$, the bosonic field can be integrated out
in the partition function defined in Eq.~(\ref{eq:gene functional}),
yielding
\begin{align}
    &Z \propto \int \mathcal{D}\lambda\, \mathcal{D}\phi\,
    e^{-S'_0[\phi,\lambda]-S_{\rm tun}[\phi]}, \\
    &S'_0[\phi,\lambda] =
    \frac{\pi\nu}{L\beta}\sum_{k \neq 0}\sum_{\omega_n}
    \frac{|e^{iks_1}-e^{iks_2}|^2}{k(vk-i\omega_n)}
    |\lambda_n|^2 \notag\\
    &\hspace{15mm}
    - \frac{i}{2\beta}\sum_{\omega_n}
    (\lambda_{-n}\phi_n + \lambda_{n}\phi_{-n}) .
\end{align}
Because the effective action $S'_0[\phi,\lambda]$ is quadratic in $\lambda_n$,
the auxiliary field $\lambda_n$ can be further integrated out, leading to
\begin{align}
    & Z \propto \int \mathcal{D}\phi\,
    e^{-S_0[\phi]-S_{\rm tun}[\phi]}, \\
    & S_0[\phi] =
    \sum_{\omega_n}
    \frac{|\phi_n|^2}{4\beta \Lambda_n} .
\end{align}
Here, the coefficient $\Lambda_n$ is given by
\begin{align}
    \Lambda_n &=
    \frac{\pi\nu}{L}\sum_{k \neq 0}
    \frac{|e^{iks_1}-e^{iks_2}|^2}{k(vk-i\omega_n)}.
\end{align}
Noting that the summation over the wavenumber is taken for
$k = 2\pi m/L$ with $m$ being a nonzero integer,
$\Lambda_n$ can be evaluated as
\begin{align}
    \Lambda_n &=
    \frac{\nu L}{\pi v} \sum_{m=1}^\infty
    \frac{1-\cos 2\pi m(s_1-s_2)/L}
    {m^2 + (\omega_n L/2\pi v)^2}.
\end{align}
Finally, by using the identity
\begin{align}
    \sum_{m=1}^\infty \frac{\cos nx}{n^2+a^2}
    = \frac{\pi}{2a}
    \frac{\cosh a(\pi-x)}{\sinh a\pi}
    - \frac{1}{2a^2},
\end{align}
valid for $0 \le x \le 2\pi$,
we arrive at the expression given in Eq.~(\ref{eq:Lambda_n}).

\section{Derivation of the density correlation function}
\label{app:density_correlator}

In this appendix, we present the derivation of the density-density correlation function.
As discussed in Sec.~\ref{subsec:formulation res}, the charge density
$\rho(s,\tau)$ can be expressed in terms of the bosonic field
$\varphi(s,\tau)$ as
\begin{align}
    \rho(s,\tau) &= \frac{1}{2\pi}\partial_s \varphi(s,\tau)
\notag \\    \label{eq:appB rho fourier}
    &= \frac{1}{L\beta}\sum_{\omega_n}\sum_{k\neq0}
    \frac{ik}{2\pi}\varphi_{k,n}e^{iks-i\omega_n\tau} .
\end{align}
Accordingly, the density-density correlation function is written as
\begin{align}
    S_{\rho\rho}(s,s',\tau) &=
    \qty(\frac{i}{2\pi L\beta})^2
    \sum_{\omega_n} \sum_{k,k'\neq 0}
    kk' \langle \varphi_{k,n} \varphi_{k',-n} \rangle \notag \\
    & \hspace{5mm} \times
    e^{iks+ik's' -i\omega_n\tau} .
\label{eq:appB Srhorho}
\end{align}
The source term associated with the external field $\eta(s,\tau)$
in the action in Eq.~(\ref{eq:S_eta}) can be rewritten, after Fourier transformation, as
\begin{align}
    \label{eq:appB S eta Fourier}
    S_{\eta}[\varphi] &=
    \frac{1}{L\beta}\sum_{\omega_n}\sum_{k\neq 0}
    \frac{ik}{2\pi}\eta_{-k,-n}\varphi_{k,n},
\end{align}
where $\eta_{-k,-n} = \eta_{k,n}^*$.
From Eqs.~(\ref{eq:gene functional with eta}) and
(\ref{eq:appB S eta Fourier}), we therefore obtain
\begin{align}
    \label{eq:appB dens corr}
    \left.
    \frac{1}{Z[0]}
    \frac{\delta^2 Z[\eta]}
    {\delta \eta_{-k,-n}\delta \eta_{-k',n}}
    \right|_{\eta = 0}
    =
    \qty(\frac{i}{2\pi L\beta})^2
    kk'
    \langle \varphi_{k,n} \varphi_{k',-n} \rangle .
\end{align}
In complete analogy with the derivation of the effective action in
Eq.~(\ref{eq:gene functional}), the fields $\varphi(s,\tau)$ and
$\lambda(\tau)$ in the generating functional in Eq.~(\ref{eq:gene functional with eta}) can be integrated out by Gaussian
integration, leaving only the local field degree of freedom
$\phi(\tau)$.
At this stage, the generating functional takes the form
\begin{align}
    \label{eq:appB gene functional with eta}
    Z[\eta] \propto
    \int \mathcal{D}\phi\,
    e^{-S_{\rm eff}[\phi]
    -S_{\eta,1}[\phi,\eta]
    -S_{\eta,2}[\eta]},
\end{align}
where
\begin{align}
    \label{eq:appB z ex1}
    & S_{\eta,1}[\phi,\eta] =
    -\sum_{\omega_n}\sum_{k\neq0}
    \frac{\nu (e^{-iks_2} - e^{-iks_1})}
    {2i \beta L \Lambda_n (vk-i\omega_n)}
    \eta_{-k,-n}\phi_n , \\
    \label{eq:appB z ex2}
    & S_{\eta,2}[\eta] =
    - \frac{\nu }{4\pi\beta L}
    \sum_{\omega_n}
    \Bigg(
    \sum_{k\neq0}
    \frac{k}{vk-i\omega_n}
    |\eta_{k,n}|^2  \notag \\
    & \hspace{5mm}
    - \sum_{k,k'\neq0}
    \frac{\nu(e^{-iks_1}-e^{-iks_2})
    (e^{ik's_1}-e^{ik's_2})}
    { \Lambda_n(vk-i\omega_n)(vk'-i\omega_n)}
    \eta^{*}_{k,n}\eta_{k',n}
    \Bigg).
\end{align}
Using Eqs.~(\ref{eq:appB gene functional with eta})--(\ref{eq:appB z ex2}),
we obtain
\begin{align}
    \label{eq:appB diff Z}
    & \frac{1}{Z[0]}
    \left.
    \frac{\delta^{2} Z[\eta]}
    {\delta \eta_{-k,-n}\delta \eta_{-k',n}}
    \right|_{\eta=0}\notag \\
    &=
    \left\langle
    \frac{\delta^{2} e^{-S_{\eta,1}[\phi,\eta]}}
    {\delta \eta_{-k,-n}\delta \eta_{-k',n}}
    \right\rangle_{\eta=0}
    +
    \left\langle
    \frac{\delta^{2} e^{-S_{\eta,2}[\eta]}}
    {\delta \eta_{-k,-n}\delta \eta_{-k',n}}
    \right\rangle_{\eta=0} ,
\end{align}
where
$\langle \cdots \rangle
= Z[0]^{-1}\int \mathcal{D}\phi\,
(\cdots)e^{-S_{\rm eff}[\phi]}$.
Evaluating each term on the right-hand side of
Eq.~(\ref{eq:appB diff Z}), we find
\begin{align}
    \left\langle
    \frac{\delta^{2} e^{-S_{\eta,1}[\phi,\eta]}}
    {\delta \eta_{-k,-n}\delta \eta_{-k',n}}
    \right\rangle_{\eta=0}
    &=
    \frac{\langle \phi_n \phi_{-n} \rangle}
    {\beta^2 \Lambda_{n}^2}
    \tilde{A}(i\omega_n), \\
    \left\langle
    \frac{\delta^{2} e^{-S_{\eta,2}[\eta]}}
    {\delta \eta_{-k,-n}\delta \eta_{-k',n}}
    \right\rangle_{\eta=0}
    &=
    -\frac{2\tilde{A}(i\omega_n)}
    {\beta\Lambda_n}
    \notag \\
    &\quad +
    \frac{\nu k}
    {2\pi L \beta(vk-i\omega_n)}
    \delta_{k,-k'}.
\end{align}
Here, we have introduced
\begin{align}
     \tilde{A}(i\omega_n) \equiv
     \sum_{i,j=1}^2
     \frac{\nu^2 (-1)^{i+j+1}
     e^{-iks_i-ik's_j}}
     {4\beta^2 L^2 \Lambda_{n}^2
     (vk-i\omega_n)(vk'+i\omega_n)}.
\end{align}
Combining these results with
Eqs.~(\ref{eq:appB Srhorho}) and (\ref{eq:appB dens corr}),
the density-density correlation function is obtained as
\begin{align}
    \label{eq:appB dens-dens corr imag time}
    S_{\rho\rho}(s,s',\tau)
    &=
    \sum_{\omega_n}
    \frac{2e^{-i\omega_n\tau}}
    {\beta\Lambda_n}
    \Biggl[
    -1
    + \frac{\langle \phi_n \phi_{-n}\rangle}
    {2\beta \Lambda_n}
    \Biggr]
    A(s,s',i\omega_n)
    \notag \\
    &\quad +
    \frac{\nu}{2\pi\beta L}
    \sum_{\omega_n}
    \sum_{k\neq0}
    \frac{k}{vk-i\omega_n}
    e^{ik(s-s') -i\omega_n\tau} .
\end{align}
The function $A(s,s',i\omega_n)$ is identical to that defined in
Eq.~(\ref{eq:A in charge dens}).
Finally, by performing a Fourier transform with respect to imaginary time,
we recover the expressions given in
Eqs.~(\ref{eq:charge dens corr})-(\ref{eq:Phi(i omega)}).

\section{Detailed calculations for IQH droplets}
\label{app:IQHE}

\begin{figure}[tb]
\centering
\includegraphics[width=50mm]{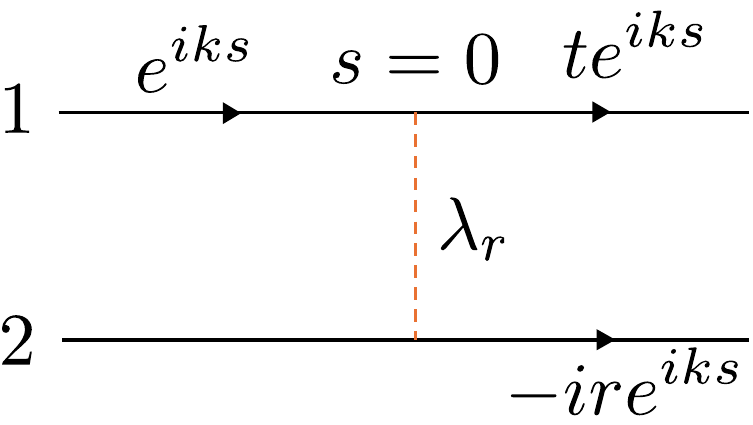}
\caption{Schematic for a wavefunction on two transmission channels for an electron incident from channel 1 toward the QPC.}
\label{fig:scattering_calc}
\end{figure}

In this appendix, we present the derivation of the results for IQH droplets given in Sec.~\ref{sec:IQHE}.
Throughout this appendix, 
we work in first quantization.
In the absence of the QPC, the edge state of the IQH droplet is described by a noninteracting chiral electron channel, which satisfies the Schr\"odinger equation 
\begin{equation}
-iv\partial_s \psi(s) = E\psi(s), \quad (0\le s < L).
\end{equation}
To relate the strength of quasiparticle tunneling $\lambda_r$ to the transmission and reflection amplitudes $t$ and $r$, we introduce $\psi_1(s) = \psi(s)$ and $\psi_2(s) = \psi(s+L/2)$, both of which are discontinuous at $s=0$ (see Fig.~\ref{fig:scattering_calc}).
In the limit $L\rightarrow \infty$, the Schr\"odinger equation in the presence of the QPC reads
\begin{align}
    \label{appE schrodinger eq}
    \left(-iv\partial_s \hat{I} + \lambda_r\delta(s) \hat{\sigma}_x\right)\left(
    \begin{matrix}
        \psi_1(s) \\
        \psi_2(s)
    \end{matrix}
    \right) = E\left(
    \begin{matrix}
        \psi_1(s) \\
        \psi_2(s)
    \end{matrix}
    \right) .
\end{align}
Here, $\hat{I}$ is the identity matrix, and $\hat{\sigma}_x$ is the $x$-component Pauli matrix. 
Consider the scattering ansatz
\begin{align}
    \label{appE psi1 def}
    \psi_1(s) &= 
    \begin{cases}
        e^{iks}, & (s<0), \\
        t e^{iks}, & (s>0), 
    \end{cases} \\
    \label{appE psi2 def}
    \psi_2(s) &= 
    \begin{cases}
        0, & (s<0), \\
        -ir e^{iks}, & (s>0). 
    \end{cases} 
\end{align}
Substituting Eqs.~(\ref{appE psi1 def}) and (\ref{appE psi2 def}) into Eq.~(\ref{appE schrodinger eq}) and integrating across $x=0$ yields
\begin{align}
    t &= \frac{1-\lambda_r^2/4v^2}{1+\lambda_r^2/4v^2}, \\
    \label{appE ref coeff}
    r &= \frac{\lambda_r/v}{1+\lambda_r^2/4v^2}.
\end{align}
Thus, $r>0$ for $\lambda_r >0$ ($\tilde{V}>0$).
In comparison with the PIMC simulation, the short-distance cutoff is identified as $\alpha = v\Delta\tau$ (see also Sec.~\ref{subsec:PIMC}).

\begin{figure}[tb]
\centering
\includegraphics[width=85mm]{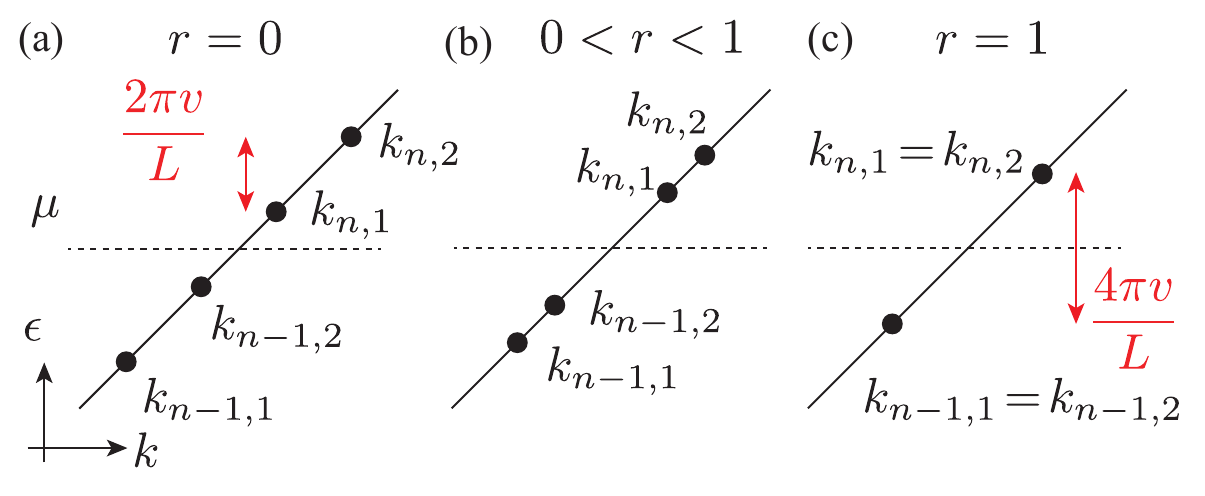}
\caption{Schematic of energy levels for (a) $r=0$ ($\chi=0$), (b) $0<r<1$ ($0<\chi<\pi/2$), and (c) $r=1$ ($\chi=\pi/2$). 
The horizontal axis represents the electron wave number, and the vertical axis the energy. 
The dashed line denotes the chemical potential.}
    \label{fig:energy_level}
\end{figure}

For a finite system, the eigenwavenumbers and eigenenergies are denoted by $k_{n,j}$ and $E_{n,j} = vk_{n,j}$, respectively.
The wavefunction of the chiral edge mode takes the form
\begin{align}
\psi(s) = \left\{ \begin{array}{ll} e^{iks+i\phi_{\rm AB} s/L}, & (0<s<L/2), \\
e^{iks+i\phi_{\rm AB} s/L+i\Theta}, & (L/2<s<L), 
\end{array} \right. 
\label{eq:wf}
\end{align}
where $\Theta$ is a phase shift at the QPC and $\phi_{\rm AB}$ is an AB phase induced by the external magnetic field.
The normalization is set to unity since the energy eigenvalues do not depend on it.
The incident and outgoing wavefunctions at the QPCs are given by
\begin{align}
\psi_{\rm in}(0) &= \psi(L-0) = e^{ikL+i\phi_{\rm AB} +i\Theta},\label{eq:pin1} \\
\psi_{\rm out}(0) &= \psi(0+0) = 1, \\
\psi_{\rm in}(L/2) &= \psi(L/2-0) = e^{ikL/2+i\phi_{\rm AB}/2}, \\
\psi_{\rm out}(L/2) &= \psi(L/2+0) = e^{ikL/2+i\phi_{\rm AB}/2+i\Theta}. \label{eq:pin4}
\end{align}
Applying the $S$ matrix in Eq.~(\ref{eq:Smatrix}), we obtain
\begin{align}
\label{eq:SmatrixEq}
\left(\begin{array}{c} \psi_{\rm out}(0) \\
\psi_{\rm out}(L/2) \end{array} \right)
= \left(
    \begin{matrix}
        t & -ir \\
        -ir & t
    \end{matrix}
    \right)
    \left(\begin{array}{c} \psi_{\rm in}(0) \\
\psi_{\rm in}(L/2) \end{array} \right) .
\end{align}
Substituting Eqs.~(\ref{eq:pin1})-(\ref{eq:pin4}) and eliminating $e^{i\Theta}$, one obtains
\begin{align}
x^2 - 2ir x - 1 = 0 ,
\label{eq:solx}
\end{align}
where $x=e^{ikL/2+i\phi_{\rm AB}/2}$.
The solutions are $x= ir \pm t = e^{i\chi}, e^{i(\pi-\chi)}$, where $\tan\chi = r/t$ ($0 \le \chi \le \pi/2$).
The wavenumbers are thus discretized as $k_{n,j} = 4\pi(n+\delta_j)/L$, where $\delta_1 = (2\chi-\phi_{\rm AB})/4\pi$ and $\delta_2 = 1/2 - (2\chi + \phi_{\rm AB})/4\pi$ denote the QPC-induced shifts.
Figure~\ref{fig:energy_level} shows a schematic energy diagram for $\phi_{\rm AB} = -\pi$.
In the absence of the QPC ($r=0$), the wavenumber reduces to $k = 2 \pi( m + 1/2)/L$ as expected.
On the other hand, for the decoupled droplets ($r=1$), the wavenumber becomes doubly degenerate, as $k = 4 \pi( n + 1/2 )/L$. 
The phase shifts follow from Eqs.~(\ref{eq:pin1})-(\ref{eq:pin4}) and are given by
\begin{align}
\Theta_j =\left\{ \begin{array}{ll} -\chi,  & (j=1), \\
\chi, & (j=2). \end{array} \right. 
\end{align} 

Using these eigenfunctions, the electron field operators are written as
\begin{align}
\label{ele ope3}
\psi(x,\tau) &= \frac{1}{\sqrt{L}}\sum_{n}\sum_{j=1,2} e^{ik_{n,j}x-vk_{n,j}\tau} g_j(x) c_{n,j}, \\
\label{ele ope2}
\psi^\dagger(x,\tau) &= \frac{1}{\sqrt{L}}\sum_{n}\sum_{j=1,2} e^{-ik_{n,j}x+vk_{n,j}\tau} g_j^*(x) c^\dagger_{n,j}, 
\end{align}
where $c_{n,j}$ ($c_{n,j}^\dagger$) denotes the annihilation (creation) operator and $g_j(x)$ is defined by
\begin{align}
g_j(x) &= \begin{cases}
1, & (x<L/2) ,\\ e^{i\Theta_j}, & (x>L/2) .
\end{cases}
\end{align}
Using the identity
\begin{align}
y(x) &= \frac{L}{2\pi} \sin \frac{2\pi x}{L} = \sum_{\eta = \pm 1} \frac{\eta L}{4\pi i} \exp\left(\frac{2\pi i \eta x}{L}\right), 
\end{align}
the imaginary-time susceptibility reads
\begin{align}
\epsilon(\tau) &= \frac{1}{L^2} \sum_{m,n} \sum_{j,j'=1}^2 e^{v(k_{n,j}-k_{m,j'})\tau} \notag \\
& \hspace{3mm} \times f(vk_{n,j})(1-f(vk_{m,j'})) I_{n,j,m,j'} I_{n,j,m,j'}^*, 
\end{align}
where $f(\epsilon)=(e^{\beta\epsilon}+1)^{-1}$ is the Fermi-Dirac distribution function and $I_{n,j,m,j'}$ is defined as
\begin{align}
I_{n,j,m,j'} &= \frac{L}{4\pi i} \sum_{\eta = \pm 1} \int_0^L dx \,   g_j^*(x) g_{j'}(x) \notag \\
& \hspace{5mm} \times e^{2\pi i\eta x/L- i
(k_{n,j}-k_{m,j'})x}.
\end{align}
Carrying out the integration yields
\begin{align}
I_{n,1,m,1} &= I_{n,2,m,2} =0 ,\\
I_{n,2,m,1} &= \frac{iL^2e^{i\chi}\sin{\chi}/4\pi^2}{(m-n+\chi/\pi)(m-n-1+\chi/\pi)}, \\
I_{n,1,m,2} &= -\frac{iL^2 e^{-i\chi} \sin \chi/4\pi^2}{(n-m+\chi/\pi)(n-m-1+\chi/\pi)}.
\end{align}
After Fourier transformation and analytic continuation ($i\omega_n \rightarrow \omega + i\delta$), we obtain Eq.~\eqref{eq:ele epsilon exact}.

\section{Detailed calculations of perturbation}
\label{app:perturbation}

Our starting point is the effective action derived in Appendix~\ref{app:effective_action}.
An auxiliary field $\xi(\tau)=\beta^{-1}\sum_n \xi_n e^{-i\omega_n\tau}$ is introduced.  
The generating functional then reads
\begin{align}
Z[\xi] &= \int \mathcal{D}\phi\, e^{-S_{\rm eff}[\phi,\xi]}, \\
S_{\rm eff}[\phi,\xi] &= S_0[\phi] + S_{\rm tun}[\phi]
+ \frac{1}{\beta}\sum_n \xi_{-n}\phi_n ,\\
S_0[\phi] &= \sum_n \frac{|\phi_n|^2}{4\beta\Lambda_n}, \\
S_{\rm tun}[\phi] &= -V \int d\tau\, \cos\phi(\tau).
\end{align}
The correlation function follows from functional differentiation
\begin{align}
\langle \phi_n \phi_{-n} \rangle
= \beta^2
\left.
\frac{\delta^2 \ln Z[\xi]}
{\delta \xi_{-n}\delta \xi_n}
\right|_{\xi=0}.
\end{align}
For $V=0$,
\begin{align}
Z_0[\xi]
= \exp\!\left(
\frac{1}{\beta}\sum_n \Lambda_n \xi_n\xi_{-n}
\right),
\end{align}
yielding the non-interacting correlator
\begin{align}
\langle \phi_n \phi_{-n} \rangle_0
= 2\beta\Lambda_n .
\end{align}
Expanding in $S_{\rm tun}$,
\begin{align}
e^{-S_0-S_{\rm tun}}
= e^{-S_0}
\left(
1 - S_{\rm tun}
+ \frac{1}{2}S_{\rm tun}^2
+ \cdots
\right),
\end{align}
the first-order contribution to the generating functional reads
\begin{align}
Z_1[\xi]
= V
\int\!\mathcal{D}\phi
\int\! d\tau\,
\cos\phi\,
e^{-S_0[\phi]
-\beta^{-1}\sum_n \xi_{-n}\phi_n}.
\end{align}
\begin{widetext}
Performing standard Gaussian integration of the $\phi$ field yields
\begin{align}
\left.
\frac{Z_1}{Z_0}
\right|_{\xi=0}
&= \beta V e^{-\bar{\Lambda}}, \\
\left.
\frac{1}{Z_0}
\frac{\delta^2 Z_1}
{\delta \xi_{-n}\delta \xi_n}
\right|_{\xi=0}
&= 2V e^{-\bar{\Lambda}}
\Lambda_n
\left(1-\frac{2\Lambda_n}{\beta}\right),
\end{align}
with $\bar{\Lambda}=\beta^{-1}\sum_p \Lambda_p$.
The first-order correction to the correlation function thus reads
\begin{align}
\langle \phi_n \phi_{-n} \rangle_1
= -4\beta V \Lambda_n^2 e^{-\bar{\Lambda}}.
\label{eq:phiphi1}
\end{align}
At second order, one has
\begin{align}
Z_2[\xi]
&= \frac{V^2}{2}
\int\!\mathcal{D}\phi
\int\! d\tau d\tau'\,
\cos\phi(\tau)\cos\phi(\tau') e^{-S_0[\phi]
-\beta^{-1}\sum_n \xi_{-n}\phi_n}.
\end{align}
Again, performing the Gaussian integration over $\phi$ gives
\begin{align}
\left.
\frac{Z_2}{Z_0}
\right|_{\xi=0}
&= +\frac{\beta V^2}{4}
e^{-2\bar{\Lambda}}
[{\cal P}_+(0)+{\cal P}_-(0)], \\
\left.
\frac{1}{Z_0}
\frac{\delta^2 Z_2}
{\delta \xi_{-n}\delta \xi_n}
\right|_{\xi=0}
&=
\frac{\Lambda_n V^2}{2}
e^{-2\bar{\Lambda}} \Biggl\{
\left(1-\frac{4\Lambda_n}{\beta}\right)
[{\cal P}_+(0)+{\cal P}_-(0)] 
+\frac{4\Lambda_n}{\beta}
[{\cal P}_+(i\omega_n)-{\cal P}_-(i\omega_n)]
\Biggr\},
\end{align}
where ${\cal P}_\pm(\omega_n)$ is defined in
Eq.~\eqref{eq:P_pm def}.
The second-order correction to the correlation function becomes
\begin{align}
\langle \phi_n \phi_{-n}\rangle_2
&= -2\beta V^2 \Lambda_n^2
e^{-2\bar{\Lambda}}
\Bigl[
{\cal P}_+(0)+{\cal P}_-(0)
-{\cal P}_+(i\omega_n)
+{\cal P}_-(i\omega_n)
-2\beta
\Bigr].
\label{eq:phiphi2}
\end{align}
For $s_1=0$ and $s_2=L/2$, the interaction kernel takes the form
\begin{align}
\Gamma(i\omega_n)
&=
\frac{8\pi^2(\omega_T^2+\omega_n^2)}
{\nu L^2}
\Biggl[
\frac{1}{\omega_T} -\frac{\omega_T^2+\omega_n^2}
{
(\omega_T+\nu\beta\Phi)(\omega_T^2+\omega_n^2)
+\nu\beta\Phi(\omega_T^2-\omega_n^2)
}
\Biggr],
\end{align}
\end{widetext}
where $\Phi (i \omega_n)$ was defined in Eq.~\eqref{eq:Phi(i omega)}.
Expanding in $V$, we are left with
\begin{align}
\Gamma(i\omega_n)
=
\left.
\frac{\partial \Gamma}{\partial V}
\right|_{V=0} V
+\frac{1}{2}
\left.
\frac{\partial^2 \Gamma}{\partial V^2}
\right|_{V=0} V^2
+\mathcal{O}(V^3).
\label{eq:expansionGamma}
\end{align}
The coefficients of the expansion are then given by
\begin{align}
\left.
\frac{\partial \Gamma}{\partial V}
\right|_{V=0}
&=
\left.
\frac{\partial \Gamma}{\partial \Phi}
\frac{\partial \Phi}{\partial V}
\right|_{V=0}, \\
\left.
\frac{\partial^2 \Gamma}{\partial V^2}
\right|_{V=0}
&=
\left.
\left(
\frac{\partial^2 \Phi}{\partial V^2}
\frac{\partial \Gamma}{\partial \Phi}
+
\left(\frac{\partial \Phi}{\partial V}\right)^2
\frac{\partial^2 \Gamma}{\partial \Phi^2}
\right)
\right|_{V=0},
\end{align}
where the required derivatives read
\begin{align}
\left.
\frac{\partial \Gamma}{\partial \Phi}
\right|_{V=0}
&=
\frac{16\pi^2\beta}{L^2}, \\
\left.
\frac{\partial^2 \Gamma}{\partial \Phi^2}
\right|_{V=0}
&=
-\frac{64\pi^2\nu\beta^2\omega_T}
{L^2(\omega_T^2+\omega_n^2)}, \\
\left.
\frac{\partial \Phi}{\partial V}
\right|_{V=0}
&=
\frac{1}{\beta^2\Lambda_n^2}
\frac{\langle \phi_n\phi_{-n}\rangle_1}{V}, \\
\left.
\frac{\partial^2 \Phi}{\partial V^2}
\right|_{V=0}
&=
\frac{2}{\beta^2\Lambda_n^2}
\frac{\langle \phi_n\phi_{-n}\rangle_2}{V^2}.
\end{align}
Substituting these back into Eq.~\eqref{eq:expansionGamma}  and using Eqs.~\eqref{eq:phiphi1} and \eqref{eq:phiphi2}, reproduces the result of Eq.~\eqref{eq:Gamma(2)}.

\bibliography{./reference}

@article{Stern2008,
  author  = {Stern, Ady},
  title   = {Anyons and the quantum Hall effect: A pedagogical review},
  journal = {Annals of Physics},
  volume  = {323},
  number  = {1},
  pages   = {204--249},
  year    = {2008},
  doi     = {10.1016/j.aop.2007.10.008}
}

@article{Bosco2019,
  author  = {Bosco, Stefano and DiVincenzo, David P. and Reilly, David J.},
  title   = {Transmission Lines and Metamaterials Based on Quantum Hall Plasmonics},
  journal = {Phys. Revi. Appl.},
  volume  = {12},
  issue   = {1},
  pages   = {014030},
  year    = {2019},
  doi     = {10.1103/PhysRevApplied.12.014030}
}

@article{Feldman2021,
  author  = {Feldman, D. E. and Halperin, B. I.},
  title   = {Fractional Charge and Fractional Statistics in the Quantum Hall Effects},
  journal = {Rep. Prog. Phys.},
  volume  = {84},
  pages   = {076501},
  year    = {2021},
  doi     = {10.1088/1361-6633/abfb15}
}

@article{Carrega2021,
  author  = {Carrega, M. and Chirolli, L. and Heun, S. and Sorba, L.},
  title   = {Anyons in Quantum Hall Interferometry},
  journal = {Nat. Rev. Phys.},
  volume  = {3},
  pages   = {698--711},
  year    = {2021},
  doi     = {10.1038/s42254-021-00351-0}
}

@article{Batra2023,
  author  = {Batra, N. and Wei, Z. and Vishveshwara, S. and Feldman, D. E.},
  title   = {Anyonic Mach--Zehnder Interferometer on a Single Edge of a Two-Dimensional Electron Gas},
  journal = {Phys. Rev. B},
  volume  = {108},
  pages   = {L241302},
  year    = {2023},
  doi     = {10.1103/PhysRevB.108.L241302}
}

@article{Ferraro2014,
  author  = {Ferraro, D. and Carrega, M. and Braggio, A. and Sassetti, M.},
  title   = {Multiple quasiparticle Hall spectroscopy
investigated with a resonant detector},
  journal = {New J. Phys.},
  volume  = {16},
  pages   = {043018},
  year    = {2014},
  doi     = {10.1088/1367-2630/16/4/043018}
}

@article{Drichko2014,
  author  = {Drichko, I. L. and Diakonov, A. M. and Malysh, V. A. and Smirnov, I. Yu. and Galperin, Y. M. and Ilyinskaya, N. D. and Usikova, A. A. and Kummer, M. and von K{\"a}nel, H.},
  title   = {Contactless measurement of alternating current conductance in quantum Hall structures},
  journal = {J. Appl. Phys.},
  volume  = {116},
  pages   = {154309},
  year    = {2014},
  doi     = {10.1063/1.4898737}
}

@article{vonDelft1998,
author = {von Delft, Jan and Schoeller, Herbert},
title = {Bosonization for beginners --- refermionization for experts},
journal = {Ann. Phys. (Berlin)},
volume = {510},
pages = {225-305},
doi = {https://doi.org/10.1002/andp.19985100401},
year = {1998}
}

@article{Halperin82,
  title = {Quantized Hall conductance, current-carrying edge states, and the existence of extended states in a two-dimensional disordered potential},
  author = {Halperin, B. I.},
  journal = {Phys. Rev. B},
  volume = {25},
  issue = {4},
  pages = {2185--2190},
  numpages = {0},
  year = {1982},
  month = {Feb},
  publisher = {American Physical Society},
  doi = {10.1103/PhysRevB.25.2185},
  url = {https://link.aps.org/doi/10.1103/PhysRevB.25.2185}
}

@article{Laughlin83,
  title = {Anomalous Quantum Hall Effect: An Incompressible Quantum Fluid with Fractionally Charged Excitations},
  author = {Laughlin, R. B.},
  journal = {Phys. Rev. Lett.},
  volume = {50},
  issue = {18},
  pages = {1395--1398},
  numpages = {0},
  year = {1983},
  month = {May},
  publisher = {American Physical Society},
  doi = {10.1103/PhysRevLett.50.1395},
  url = {https://link.aps.org/doi/10.1103/PhysRevLett.50.1395}
}

@article{Wilczek82,
  title = {Quantum Mechanics of Fractional-Spin Particles},
  author = {Wilczek, Frank},
  journal = {Phys. Rev. Lett.},
  volume = {49},
  issue = {14},
  pages = {957--959},
  numpages = {0},
  year = {1982},
  month = {Oct},
  publisher = {American Physical Society},
  doi = {10.1103/PhysRevLett.49.957},
  url = {https://link.aps.org/doi/10.1103/PhysRevLett.49.957}
}

@article{Arovas84,
  title = {Fractional Statistics and the Quantum Hall Effect},
  author = {Arovas, Daniel and Schrieffer, J. R. and Wilczek, Frank},
  journal = {Phys. Rev. Lett.},
  volume = {53},
  issue = {7},
  pages = {722--723},
  numpages = {0},
  year = {1984},
  month = {Aug},
  publisher = {American Physical Society},
  doi = {10.1103/PhysRevLett.53.722},
  url = {https://link.aps.org/doi/10.1103/PhysRevLett.53.722}
}

@article{Nayak08,
  title = {Non-Abelian anyons and topological quantum computation},
  author = {Nayak, Chetan and Simon, Steven H. and Stern, Ady and Freedman, Michael and Das Sarma, Sankar},
  journal = {Rev. Mod. Phys.},
  volume = {80},
  issue = {3},
  pages = {1083--1159},
  numpages = {0},
  year = {2008},
  month = {Sep},
  publisher = {American Physical Society},
  doi = {10.1103/RevModPhys.80.1083},
  url = {https://link.aps.org/doi/10.1103/RevModPhys.80.1083}
}

@article{Saleur02,
title = {Edge states tunneling in the fractional quantum Hall effect: integrability and transport},
journal = {Comptes Rendus Physique},
volume = {3},
number = {6},
pages = {685-695},
year = {2002},
issn = {1631-0705},
doi = {https://doi.org/10.1016/S1631-0705(02)01366-X},
url = {https://www.sciencedirect.com/science/article/pii/S163107050201366X},
author = {Hubert Saleur}
}

@Inbook{Glattli05,
author="Glattli, D. Christian",
editor="Dou{\c{c}}ot, Beno{\^i}t
and Pasquier, Vincent
and Duplantier, Bertrand
and Rivasseau, Vincent",
title="Tunneling Experiments in the Fractional Quantum Hall Effect Regime",
bookTitle="The Quantum Hall Effect: Poincar{\'e} Seminar 2004",
year="2005",
publisher="Birkh{\"a}user Basel",
address="Basel",
pages="163--197",
isbn="978-3-7643-7393-1",
doi="10.1007/3-7643-7393-8_5",
url="https://doi.org/10.1007/3-7643-7393-8_5"
}

@article{Kane95,
  title = {Contacts and edge-state equilibration in the fractional quantum Hall effect},
  author = {Kane, C. L. and Fisher, Matthew P. A.},
  journal = {Phys. Rev. B},
  volume = {52},
  issue = {24},
  pages = {17393--17405},
  numpages = {0},
  year = {1995},
  month = {Dec},
  publisher = {American Physical Society},
  doi = {10.1103/PhysRevB.52.17393},
  url = {https://link.aps.org/doi/10.1103/PhysRevB.52.17393}
}

@article{Deviatov06,
  title = {Equilibration between edge states in the fractional quantum Hall effect regime at high imbalances},
  author = {Deviatov, E. V. and Kapustin, A. A. and Dolgopolov, V. T. and Lorke, A. and Reuter, D. and Wieck, A. D.},
  journal = {Phys. Rev. B},
  volume = {74},
  issue = {7},
  pages = {073303},
  numpages = {4},
  year = {2006},
  month = {Aug},
  publisher = {American Physical Society},
  doi = {10.1103/PhysRevB.74.073303},
  url = {https://link.aps.org/doi/10.1103/PhysRevB.74.073303}
}

@article{Mast85,
  title = {{Observation of Bulk and Edge Magnetoplasmons in a Two-Dimensional Electron Fluid}},
  author = {Mast, D. B. and Dahm, A. J. and Fetter, A. L.},
  journal = {Phys. Rev. Lett.},
  volume = {54},
  issue = {15},
  pages = {1706--1709},
  numpages = {0},
  year = {1985},
  month = {Apr},
  publisher = {American Physical Society},
  doi = {10.1103/PhysRevLett.54.1706},
  url = {https://link.aps.org/doi/10.1103/PhysRevLett.54.1706}
}

@article{Glattli85,
  title = {{Dynamical Hall Effect in a Two-Dimensional Classical Plasma}},
  author = {Glattli, D. C. and Andrei, E. Y. and Deville, G. and Poitrenaud, J. and Williams, F. I. B.},
  journal = {Phys. Rev. Lett.},
  volume = {54},
  issue = {15},
  pages = {1710--1713},
  numpages = {0},
  year = {1985},
  month = {Apr},
  publisher = {American Physical Society},
  doi = {10.1103/PhysRevLett.54.1710},
  url = {https://link.aps.org/doi/10.1103/PhysRevLett.54.1710}
}

@article{Magnus54,
author = {Magnus, Wilhelm},
title = {On the exponential solution of differential equations for a linear operator},
journal = {Commun. Pure Appl. Math.},
volume = {7},
number = {4},
pages = {649-673},
doi = {https://doi.org/10.1002/cpa.3160070404},
url = {https://onlinelibrary.wiley.com/doi/abs/10.1002/cpa.3160070404},
year = {1954}
}

@article{Cano2013PRB,
  title = {{Microwave absorption by a mesoscopic quantum Hall droplet}},
  author = {{J. Cano, A. Doherty, C. Nayak, D. Reilly}},
  journal = {Phys. Rev. B},
  volume = {88},
  pages = {165305},
  year = {2013},
  doi = {10.1103/PhysRevB.88.165305},
  url = {https://link.aps.org/doi/10.1103/PhysRevB.88.165305}
}

@article{Moon1993PRL,
  title = {{Resonant Tunneling between Quantum Hall Edge States}},
  author = {{K. Moon, H. Yi, C. L. Kane, S. M. Girvin, and M. P. A. Fisher}},
  journal = {Phys. Rev. Lett.},
  volume = {71},
  pages = {4381},
  year = {1993},
  doi = {10.1103/PhysRevLett.71.4381},
}

@article{Michael1997PRB,
  title = {{Aharanov-Bohm effect in the chiral Luttinger liquid}},
  author = {{M. R. Geller, D. Loss}},
  journal = {Phys. Rev. B},
  volume = {56},
  pages = {9692},
  year = {1997},
  doi = {10.1103/PhysRevB.56.9692},
}

@article{Loss1992PRL,
  title = {{Parity effects in a Luttinger liquid: Diamagnetic and paramagnetic ground states}},
  author = {Loss, Daniel},
  journal = {Phys. Rev. Lett.},
  volume = {69},
  issue = {2},
  pages = {343--346},
  numpages = {0},
  year = {1992},
  month = {Jul},
  publisher = {American Physical Society},
  doi = {10.1103/PhysRevLett.69.343},
  url = {https://link.aps.org/doi/10.1103/PhysRevLett.69.343}
}

@article{Jonckheere2023PRL,
  title = {{Anyonic Statistics Revealed by the Hong-Ou-Mandel Dip for Fractional Excitations}},
  author = {Jonckheere, T. and Rech, J. and Gr\'emaud, B. and Martin, T.},
  journal = {Phys. Rev. Lett.},
  volume = {130},
  issue = {18},
  pages = {186203},
  numpages = {6},
  year = {2023},
  month = {May},
  publisher = {American Physical Society},
  doi = {10.1103/PhysRevLett.130.186203},
  url = {https://link.aps.org/doi/10.1103/PhysRevLett.130.186203}
}

@article{Schiller2023PRL,
  title = {{Anyon Statistics through Conductance Measurements of Time-Domain Interferometry}},
  author = {Schiller, Noam and Shapira, Yotam and Stern, Ady and Oreg, Yuval},
  journal = {Phys. Rev. Lett.},
  volume = {131},
  issue = {18},
  pages = {186601},
  numpages = {7},
  year = {2023},
  month = {Nov},
  publisher = {American Physical Society},
  doi = {10.1103/PhysRevLett.131.186601},
  url = {https://link.aps.org/doi/10.1103/PhysRevLett.131.186601}
}

@article{Morel2022PRB,
  title = {Fractionalization and anyonic statistics in the integer quantum Hall collider},
  author = {Morel, Tom and Lee, June-Young M. and Sim, H.-S. and Mora, Christophe},
  journal = {Phys. Rev. B},
  volume = {105},
  issue = {7},
  pages = {075433},
  numpages = {14},
  year = {2022},
  month = {Feb},
  publisher = {American Physical Society},
  doi = {10.1103/PhysRevB.105.075433},
  url = {https://link.aps.org/doi/10.1103/PhysRevB.105.075433}
}

@article{Lee2022Nat,
	author = {Lee, June-Young M. and Sim, H. -S.},
	doi = {10.1038/s41467-022-34329-y},
	journal = {Nat. Commun.},
	pages = {6660},
	title = {{Non-Abelian anyon collider}},
	volume = {13},
	year = {2022},
}

@article{Willett2023PRX,
  title = {{Interference Measurements of Non-Abelian $e/4$ \& Abelian $e/2$ Quasiparticle Braiding}},
  author = {Willett, R. L. and Shtengel, K. and Nayak, C. and Pfeiffer, L. N. and Chung, Y. J. and Peabody, M. L. and Baldwin, K. W. and West, K. W.},
  journal = {Phys. Rev. X},
  volume = {13},
  issue = {1},
  pages = {011028},
  numpages = {19},
  year = {2023},
  month = {Mar},
  publisher = {American Physical Society},
  doi = {10.1103/PhysRevX.13.011028},
  url = {https://link.aps.org/doi/10.1103/PhysRevX.13.011028}
}

@article{Nakamura2020,
	author = {Nakamura, J. and Liang, S. and Gardner, G. C. and Manfra, M. J.},
	doi = {10.1038/s41567-020-1019-1},
	journal = {Nat. Phys.},
	pages = {931--936},
	title = {Direct observation of anyonic braiding statistics},
	volume = {16},
	year = {2020},
}

@article{Nakamura2023PRX,
  title = {{Fabry-P\'erot Interferometry at the $\ensuremath{\nu}=2/5$ Fractional Quantum Hall State}},
  author = {Nakamura, J. and Liang, S. and Gardner, G. C. and Manfra, M. J.},
  journal = {Phys. Rev. X},
  volume = {13},
  issue = {4},
  pages = {041012},
  numpages = {11},
  year = {2023},
  month = {Oct},
  publisher = {American Physical Society},
  doi = {10.1103/PhysRevX.13.041012},
  url = {https://link.aps.org/doi/10.1103/PhysRevX.13.041012}
}

@article{Glidic2023PRX,
  title = {{Cross-Correlation Investigation of Anyon Statistics in the $\ensuremath{\nu}=1/3$ and $2/5$ Fractional Quantum Hall States}},
  author = {Glidic, P. and Maillet, O. and Aassime, A. and Piquard, C. and Cavanna, A. and Gennser, U. and Jin, Y. and Anthore, A. and Pierre, F.},
  journal = {Phys. Rev. X},
  volume = {13},
  issue = {1},
  pages = {011030},
  numpages = {19},
  year = {2023},
  month = {Mar},
  publisher = {American Physical Society},
  doi = {10.1103/PhysRevX.13.011030},
  url = {https://link.aps.org/doi/10.1103/PhysRevX.13.011030}
}

@article{Lee2023Nat,
	author = {Lee, June-Young M. and Hong, Changki and Alkalay, Tomer and Schiller, Noam and Umansky, Vladimir and Heiblum, Moty and Oreg, Yuval and Sim, H. -S.},
	doi = {10.1038/s41586-023-05883-2},
	journal = {Nature},
	number = {7960},
	pages = {277--281},
	title = {Partitioning of diluted anyons reveals their braiding statistics},
	volume = {617},
	year = {2023},
	}

@article{Kundu2023Nat,
author = {{H. K. Kundu, S. Biswas, N. Ofek, V. Umansky, and M. Heiblum}},
doi = {10.1038/s41586-023-05883-2},
journal = {Nat. Phys.},
pages = {515--521},
title = {{Anyonic interference and braiding phase in a Mach-Zehnder interferometer}},
volume = {19},
year = {2023},
}

@article{Ruelle2023PRX,
  title = {{Comparing Fractional Quantum Hall Laughlin and Jain Topological Orders with the Anyon Collider}},
  author = {Ruelle, M. and Frigerio, E. and Berroir, J.-M. and Pla\c{c}ais, B. and Rech, J. and Cavanna, A. and Gennser, U. and Jin, Y. and F\`eve, G.},
  journal = {Phys. Rev. X},
  volume = {13},
  issue = {1},
  pages = {011031},
  numpages = {18},
  year = {2023},
  month = {Mar},
  publisher = {American Physical Society},
  doi = {10.1103/PhysRevX.13.011031},
  url = {https://link.aps.org/doi/10.1103/PhysRevX.13.011031}
}

@article{dePicciotto1997Nat,
	author = {de-Picciotto, R. and Reznikov, M. and Heiblum, M. and Umansky, V. and Bunin, G. and Mahalu, D.},
	doi = {10.1038/38241},
	journal = {Nature},
	pages = {162--164},
	title = {Direct observation of a fractional charge},
	volume = {389},
	year = {1997}}

@article{Bartolomei2020Sci,
    author = {H. Bartolomei  and M. Kumar  and R. Bisognin  and A. Marguerite  and J.-M. Berroir  and E. Bocquillon  and B. Pla\c{c}ais  and A. Cavanna  and Q. Dong  and U. Gennser  and Y. Jin  and G. F\`eve },
    title = {Fractional statistics in anyon collisions},
    journal = {Science},
    volume = {368},
    number = {6487},
    pages = {173-177},
    year = {2020},
    doi = {10.1126/science.aaz5601},
    URL = {https://www.science.org/doi/abs/10.1126/science.aaz5601}
}

@article{Tsui1982PRL,
  title = {Two-Dimensional Magnetotransport in the Extreme Quantum Limit},
  author = {Tsui, D. C. and Stormer, H. L. and Gossard, A. C.},
  journal = {Phys. Rev. Lett.},
  volume = {48},
  issue = {22},
  pages = {1559--1562},
  numpages = {0},
  year = {1982},
  month = {May},
  publisher = {American Physical Society},
  doi = {10.1103/PhysRevLett.48.1559},
  url = {https://link.aps.org/doi/10.1103/PhysRevLett.48.1559}
}

@article{Flavio2025PRB,
  title = {{Probing anyon statistics on a single-edge loop in the fractional quantum Hall regime}},
  author = {Ronetti, Flavio and Demazure, No\'e and Rech, J\'er\^ome and Jonckheere, Thibaut and Gr\'emaud, Beno\^{\i}t and Raymond, Laurent and Hashisaka, Masayuki and Kato, Takeo and Martin, Thierry},
  journal = {Phys. Rev. B},
  volume = {112},
  issue = {12},
  pages = {125166},
  numpages = {17},
  year = {2025},
  month = {Sep},
  publisher = {American Physical Society},
  doi = {10.1103/tdkj-k211},
  url = {https://link.aps.org/doi/10.1103/tdkj-k211}
}

@article{Chang2003RevModPhys,
  title = {{Chiral Luttinger liquids at the fractional quantum Hall edge}},
  author = {Chang, A. M.},
  journal = {Rev. Mod. Phys.},
  volume = {75},
  issue = {4},
  pages = {1449--1505},
  numpages = {0},
  year = {2003},
  month = {Nov},
  publisher = {American Physical Society},
  doi = {10.1103/RevModPhys.75.1449},
  url = {https://link.aps.org/doi/10.1103/RevModPhys.75.1449}
}

@article{Wen1992IJMP,
    author = {Wen, Xiao-Gang},
    title = {{Theory of the Edge States in Fractional Quantum Hall Effects}},
    journal = {Int. J. Mod. Phys. B},
    volume = {06},
    number = {10},
    pages = {1711-1762},
    year = {1992},
    doi = {10.1142/S0217979292000840},
}

@article{Saminadayar1997PRL,
  title = {{Observation of the $\mathit{e}\mathit{/}3$ Fractionally Charged Laughlin Quasiparticle}},
  author = {Saminadayar, L. and Glattli, D. C. and Jin, Y. and Etienne, B.},
  journal = {Phys. Rev. Lett.},
  volume = {79},
  issue = {13},
  pages = {2526--2529},
  numpages = {0},
  year = {1997},
  month = {Sep},
  publisher = {American Physical Society},
  doi = {10.1103/PhysRevLett.79.2526},
}

@article{Frigerio2024CommPhys,
	author = {Frigerio, Elric and Rebora, Giacomo and Ruelle, M{\'e}lanie and Souquet-Basi{\`e}ge, Hubert and Jin, Yong and Gennser, Ulf and Cavanna, Antonella and Pla{\c c}ais, Bernard and Baudin, Emmanuel and Berroir, Jean-Marc and Safi, In{\`e}s and Degiovanni, Pascal and F{\`e}ve, Gwendal and M{\'e}nard, Gerbold C.},
	journal = {Commun. Phys.},
	number = {1},
	pages = {314},
	title = {Gate tunable edge magnetoplasmon resonators},
	volume = {7},
	year = {2024},
    doi={10.1038/s42005-024-01803-6}
}

@article{Mahoney2017PRX,
  title = {On-Chip Microwave Quantum Hall Circulator},
  author = {Mahoney, A. C. and Colless, J. I. and Pauka, S. J. and Hornibrook, J. M. and Watson, J. D. and Gardner, G. C. and Manfra, M. J. and Doherty, A. C. and Reilly, D. J.},
  journal = {Phys. Rev. X},
  volume = {7},
  issue = {1},
  pages = {011007},
  numpages = {9},
  year = {2017},
  month = {Jan},
  publisher = {American Physical Society},
  doi = {10.1103/PhysRevX.7.011007},
  url = {https://link.aps.org/doi/10.1103/PhysRevX.7.011007}
}

@article{Oblak2024PRX,
  title = {{Anisotropic Quantum Hall Droplets}},
  author = {Oblak, Blagoje and Lapierre, Bastien and Moosavi, Per and St\'ephan, Jean-Marie and Estienne, Benoit},
  journal = {Phys. Rev. X},
  volume = {14},
  issue = {1},
  pages = {011030},
  numpages = {32},
  year = {2024},
  month = {Feb},
  publisher = {American Physical Society},
  doi = {10.1103/PhysRevX.14.011030},
  url = {https://link.aps.org/doi/10.1103/PhysRevX.14.011030}
}

@article{Rosenow2016PRL,
  title = {Current Correlations from a Mesoscopic Anyon Collider},
  author = {Rosenow, Bernd and Levkivskyi, Ivan P. and Halperin, Bertrand I.},
  journal = {Phys. Rev. Lett.},
  volume = {116},
  issue = {15},
  pages = {156802},
  numpages = {5},
  year = {2016},
  month = {Apr},
  publisher = {American Physical Society},
  doi = {10.1103/PhysRevLett.116.156802},
  url = {https://link.aps.org/doi/10.1103/PhysRevLett.116.156802}
}

@article{Han2016NatComm,
	author = {Han, Cheolhee and Park, Jinhong and Gefen, Yuval and Sim, H. -S.},
	doi = {10.1038/ncomms11131},
	journal = {Nat. Commun.},
	pages = {11131},
	title = {Topological vacuum bubbles by anyon braiding},
	volume = {7},
	year = {2016},
}

@misc{Mora2022arXiv,
      title={Anyonic exchange in a beam splitter}, 
      author={Christophe Mora},
      year={2022},
      eprint={2212.05123},
      archivePrefix={arXiv},
}

@article{Iyer2024PRL,
  title = {Finite Width of Anyons Changes Their Braiding Signature},
  author = {Iyer, K. and Ronetti, F. and Gr\'emaud, B. and Martin, T. and Rech, J. and Jonckheere, T.},
  journal = {Phys. Rev. Lett.},
  volume = {132},
  issue = {21},
  pages = {216601},
  numpages = {6},
  year = {2024},
  month = {May},
  publisher = {American Physical Society},
  doi = {10.1103/PhysRevLett.132.216601},
  url = {https://link.aps.org/doi/10.1103/PhysRevLett.132.216601}
}

@article{Thamm2024PRL,
  title = {Effect of the Soliton Width on Nonequilibrium Exchange Phases of Anyons},
  author = {Thamm, Matthias and Rosenow, Bernd},
  journal = {Phys. Rev. Lett.},
  volume = {132},
  issue = {15},
  pages = {156501},
  numpages = {6},
  year = {2024},
  month = {Apr},
  publisher = {American Physical Society},
  doi = {10.1103/PhysRevLett.132.156501},
  url = {https://link.aps.org/doi/10.1103/PhysRevLett.132.156501}
}

@article{Werkmeister2024NatComm,
	author = {Werkmeister, Thomas and Ehrets, James R. and Ronen, Yuval and Wesson, Marie E. and Najafabadi, Danial and Wei, Zezhu and Watanabe, Kenji and Taniguchi, Takashi and Feldman, D. E. and Halperin, Bertrand I. and Yacoby, Amir and Kim, Philip},
	doi = {10.1038/s41467-024-50695-1},
	journal = {Nat. Commun.},
	pages = {6533},
	title = {{Strongly coupled edge states in a graphene quantum Hall interferometer}},
	volume = {15},
	year = {2024},
}

@article{Werkmeister2025Science,
author = {Thomas Werkmeister  and James R. Ehrets  and Marie E. Wesson  and Danial H. Najafabadi  and Kenji Watanabe  and Takashi Taniguchi  and Bertrand I. Halperin  and Amir Yacoby  and Philip Kim },
title = {Anyon braiding and telegraph noise in a graphene interferometer},
journal = {Science},
volume = {388},
number = {6748},
pages = {730-735},
year = {2025},
doi = {10.1126/science.adp5015},
}

@article{Ronetti2025PRL,
  title = {Anyon Braiding on the Single Edge of a Fractional Quantum Hall State},
  author = {Ronetti, Flavio and Demazure, No\'e and Rech, J\'er\^ome and Jonckheere, Thibaut and Gr\'emaud, Beno\^{\i}t and Raymond, Laurent and Hashisaka, Masayuki and Kato, Takeo and Martin, Thierry},
  journal = {Phys. Rev. Lett.},
  volume = {135},
  issue = {14},
  pages = {146601},
  numpages = {6},
  year = {2025},
  month = {Sep},
  publisher = {American Physical Society},
  doi = {10.1103/yq1x-kxgm},
  url = {https://link.aps.org/doi/10.1103/yq1x-kxgm}
}

@article{Ruelle2025Science,
author = {M. Ruelle  and E. Frigerio  and E. Baudin  and J.-M. Berroir  and B. Pla\c{c}ais  and B. Gr\'{e}maud  and T. Jonckheere  and T. Martin  and J. Rech  and A. Cavanna  and U. Gennser  and Y. Jin  and G. M\'{e}nard  and G. F\`{e}ve },
title = {Time-domain braiding of anyons},
journal = {Science},
volume = {389},
number = {6755},
pages = {eadm7695},
year = {2025},
doi = {10.1126/science.adm7695},
}

@article{Volkov1988JETP,
  title={Edge magnetoplasmons: low frequency weakly damped excitations in inhomogeneous two-dimensional electron systems},
  author={Volkov, VA and Mikhailov, Sergey A},
  journal={Sov. Phys. JETP},
  volume={67},
  number={8},
  pages={1639--1653},
  year={1988}
}

@article{Andrei1988SurfSci,
  title={Low frequency collective excitations in the quantum-hall system},
  author={Andrei, EY and Glattli, DC and Williams, FIB and Heiblum, M},
  journal={Surf. Sci.},
  volume={196},
  number={1-3},
  pages={501--506},
  year={1988},
  doi={10.1016/0039-6028(88)90732-7},
}

@article{Talyanskii1990SurfSci,
title = {{Edge magnetoplasmons in the quantum Hall effect regime}},
journal = {Surf. Sci.},
volume = {229},
number = {1},
pages = {40-42},
year = {1990},
issn = {0039-6028},
doi = {https://doi.org/10.1016/0039-6028(90)90827-U},
author = {V.K. Talyanskii and M. Wassermeier and A. Wixforth and J. Oshinowo and J.P. Kotthaus and I.E. Batov and G. Weimann and H. Nickel and W. Schlapp},
}

@article{Ashoori1994PRB,
  title = {Edge magnetoplasmons in the time domain},
  author = {Ashoori, R. C. and Stormer, H. L. and Pfeiffer, L. N. and Baldwin, K. W. and West, K.},
  journal = {Phys. Rev. B},
  volume = {45},
  issue = {7},
  pages = {3894--3897},
  numpages = {0},
  year = {1992},
  month = {Feb},
  publisher = {American Physical Society},
  doi = {10.1103/PhysRevB.45.3894},
  url = {https://link.aps.org/doi/10.1103/PhysRevB.45.3894}
}

@article{Zhitenev1994PRB,
  title = {Experimental determination of the dispersion of edge magnetoplasmons confined in edge channels},
  author = {Zhitenev, N. B. and Haug, R. J. and Klitzing, K. v. and Eberl, K.},
  journal = {Phys. Rev. B},
  volume = {49},
  issue = {11},
  pages = {7809--7812},
  numpages = {0},
  year = {1994},
  month = {Mar},
  publisher = {American Physical Society},
  doi = {10.1103/PhysRevB.49.7809},
  url = {https://link.aps.org/doi/10.1103/PhysRevB.49.7809}
}

@article{Kumada2011PRB,
  title={Edge magnetoplasmon transport in gated and ungated quantum Hall systems},
  author={Kumada, N and Kamata, H and Fujisawa, T},
  journal={Physical Review B—Condensed Matter and Materials Physics},
  volume={84},
  number={4},
  pages={045314},
  year={2011},
  publisher={APS}
}

@article{Petkovic2013PRL,
  title={{Carrier Drift Velocity and Edge Magnetoplasmons in Graphene}},
  author={Petkovi{\'c}, I and Williams, FIB and Bennaceur, Keyan and Portier, Fabien and Roche, Patrice and Glattli, D Christian},
  journal={Phys. Rev. Lett.},
  volume={110},
  pages={016801},
  year={2013},
  doi={10.1103/PhysRevLett.110.016801}
}

@article{Kumada2014PRL,
  title={{Resonant Edge Magnetoplasmons and Their Decay in Graphene}},
  author={Kumada, Norio and Roulleau, P and Roche, B and Hashisaka, M and Hibino, H and Petkovi{\'c}, I and Glattli, DC},
  journal={Phys. Rev. Lett.},
  volume={113},
  number={26},
  pages={266601},
  year={2014},
  doi={10.1103/PhysRevLett.113.266601}
}

@article{Hashisaka2013PRB,
  title = {Distributed-element circuit model of edge magnetoplasmon transport},
  author = {Hashisaka, Masayuki and Kamata, Hiroshi and Kumada, Norio and Washio, Kazuhisa and Murata, Ryuji and Muraki, Koji and Fujisawa, Toshimasa},
  journal = {Phys. Rev. B},
  volume = {88},
  issue = {23},
  pages = {235409},
  numpages = {12},
  year = {2013},
  month = {Dec},
  publisher = {American Physical Society},
  doi = {10.1103/PhysRevB.88.235409},
  url = {https://link.aps.org/doi/10.1103/PhysRevB.88.235409}
}

@article{Endo2018JPSJ,
author = {Endo ,Akira and Koike ,Keita and Katsumoto ,Shingo and Iye ,Yasuhiro},
title = {{Frequencies of the Edge-Magnetoplasmon Excitations in Gated Quantum Hall Edges}},
journal = {J. Phys. Soc. Jpn.},
volume = {87},
number = {6},
pages = {064709},
year = {2018},
doi = {10.7566/JPSJ.87.064709},
}

@dataset{Murabayashi2026zenodo,
  author = {Murabayashi, Fumihiro},
  title  = {Code and data for {\textquotedblleft}{M}icrowave response of
            fractional quantum Hall droplets with
            quasiparticle tunneling{\textquotedblright}
            },
  note = {{Z}enodo, 2026, \url{https://doi.org/10.5281/zenodo.18933586}},
  month = mar,
}
\clearpage

\end{document}